\documentclass[
 amssymb,superscriptaddress,twocolumn]{revtex4-2}

\usepackage{amsthm}
\usepackage{mathtools}
\usepackage{graphicx}
\usepackage{xcolor}

\usepackage{braket}
\usepackage{amsmath}
\usepackage{leftindex,tensor}

\usepackage[colorlinks,citecolor=teal,linkcolor=teal,urlcolor=teal]{hyperref}

\usepackage{physics}
\usepackage{times}
\graphicspath{ {.} }
\usepackage[normalem]{ulem}

\let\oldsqrt\sqrt
\def\sqrt{\mathpalette\DHLhksqrt} \def\DHLhksqrt#1#2{%
\setbox0=\hbox{$#1\oldsqrt{#2\,}$}\dimen0=\ht0
\advance\dimen0-0.2\ht0
\setbox2=\hbox{\vrule height\ht0 depth -\dimen0}%
{\box0\lower0.4pt\box2}}

\def\b0{\boldsymbol{0}}

\newcommand{\R}     {\mathbb{R}}

\newcommand{\Ccal}   {{\mathcal C }}

\newcommand{\Fcal}   {{\mathcal F }}

\newcommand{\ph}{\phantom{}}
\newcommand{\phd}{{\vphantom{\dagger}}}
\newcommand{\veps}{\varepsilon}

\definecolor{darkgreen}{rgb}{0.0, 0.42, 0.24}

\begin{document}

\title{Optimal stellar rank approximation of squeezed cat states with photon catalysis} 

\author{Julian K. Nauth} 

\affiliation{Dahlem Center for Complex Quantum Systems, Freie Universität Berlin, 14195 Berlin, Germany}

\author{Nathan Walk}
\affiliation{Dahlem Center for Complex Quantum Systems, Freie Universität Berlin, 14195 Berlin, Germany}

\author{Ananga M. Datta}
\affiliation{Humboldt-Universit\"at zu Berlin, Institut f\"ur Physik, Institut für Physik, Newtonstr. 15, 12489 Berlin, Germany}
\affiliation{Okinawa Institute of Science and Technology Graduate University, Onna-son, Okinawa 904-0495, Japan}

\author{Kurt Busch}
\affiliation{Humboldt-Universit\"at zu Berlin, Institut f\"ur Physik, Institut für Physik, Newtonstr. 15, 12489 Berlin, Germany}
\affiliation{Max-Born-Institut, Max-Born-Str. 2A, 12489 Berlin, Germany}

\author{Jens Eisert}
\affiliation{Dahlem Center for Complex Quantum Systems, Freie Universität Berlin, 14195 Berlin, Germany}
\affiliation{Helmholtz-Zentrum Berlin für Materialien und Energie, 14109 Berlin, Germany}

\author{Oliver Benson}
\affiliation{Humboldt-Universit\"at zu Berlin, Institut f\"ur Physik, Institut für Physik, Newtonstr. 15, 12489 Berlin, Germany}

\author{Roger A. Kögler}
\affiliation{Humboldt-Universit\"at zu Berlin, Institut f\"ur Physik, Institut für Physik, Newtonstr. 15, 12489 Berlin, Germany}

\begin{abstract}
Non-Gaussian quantum states and operations constitute essential resources for achieving quantum computational advantage and enabling quantum error correction in bosonic platforms. However, their generation in optical settings remains a challenging experimental task, often relying on probabilistic heralded protocols. Here, we present an in-depth analysis of the suitability of photon catalysis between low number Fock states and squeezed states for the generation of squeezed coherent state superpositions. We employ the stellar rank formalism to characterize the non-Gaussian complexity of input resources (including both states and measurements) and the generated states. This enables a systematic comparison of the fidelity between the catalyzed output and the target states to the maximum fidelity achievable by any protocol with the same non-Gaussian input resources. In this sense, we identify instances where the catalysis protocols considered here are provably optimal. We identify parameter regimes in which high-fidelity approximations of the target states can be achieved with minimal resources. Furthermore, we benchmark the performance of photon catalysis against Gaussian boson sampling–inspired protocols in terms of success probability and state quality, highlighting the advantages of deterministic Fock state sources. We also investigate the generation of related non-Gaussian resources including squeezed Fock states, relevant for quantum error correction. To account for experimental imperfections, we model losses across all optical modes using a Hilbert space truncation approach in the Fock basis and analyze the robustness of the generated states under realistic conditions. Our results quantify the trade-offs between non-Gaussian resource complexity, achievable fidelity, and losses in photon catalysis protocols, providing practical guidelines for near-term photonic implementations.
%

\end{abstract}

\maketitle
\section{Introduction}

Among the possible hardware platforms of 
quantum computing, optical bosonic platforms take a special role in that, once certain bottlenecks have been overcome, 
they promise to be particularly scalable \cite{PsiQuantumPlatform,Xanadu,FusionBased}. 
In one way or another, 
photonic bosonic quantum computing will have to rely on some instance of non-Gaussian quantum states, where Wigner negativity is a necessary condition for achieving quantum advantage \cite{mari2012positive}. This applies to schemes of computation as such,
but prominently also to code words in notions of quantum error correction, 
where families of states such as \emph{Gottesman-Kitaev-Preskill} (GKP) states \cite{gottesman2001encoding,Lattices,ReviewGKP} and Schrödinger cat states \cite{yurke1986generating,Jeong:2002,Ralph:2003}
are prominently discussed. Despite their 
key technological importance, and even in the light of substantial progress along these lines 
\cite{cai2021bosonic}, 
the generation of such states remains experimentally challenging, but impressive progress is currently being made.
This applies to hardware-efficient quantum error correction via concatenated bosonic qubits
\cite{Putterman} as well as to GKP quantum error correction achieving the break-even point
 \cite{GKPBreakEven}, albeit in superconducting platforms.

The deterministic generation of non-Gaussian states is {challenging since}  they rely on Hamiltonians that are at least cubic in their canonical operators \cite{serafini2023quantum,walschaers2021non}. In practice, strong non-linear interactions of third order or higher are required, which {is demanding given the current state of technology} \cite{zhang2020high,fukui2022generating}. Consequently, probabilistic protocols have become the primary approach for non-Gaussian 
quantum state engineering \cite{walschaers2021non,furusawa2015quantum}. A prominent approach in photonic platforms relies on \emph{Gaussian-Boson-sampling}-like (GBS-like)  \cite{PhysRevLett.113.100502,PhysRevLett.119.170501,SupremacyReview} methods, where Gaussian states are entangled via linear optical operations and conditional non-Gaussian measurements such as \emph{photon-number-resolving detection} (PNRD) \cite{dakna1997generating,dakna1998quantum,ourjoumtsev2006generating,wakui2007photon,marek2008generating,zavatta2008subtracting,gerrits2010generation,bimbard2010quantum,neergaard2010optical,namekata2010non,yukawa2013generating,dong2014generation,takase2021generation,gorshenin2025preparation,larsen2025integrated}. Iterative applications of such schemes, where previously generated non-Gaussian output states can be reused, can enhance the amount of non-Gaussianity. This favors considering weakly non-Gaussian states as input states. In particular, Fock state vectors $\ket{n}$ with low photon number $n$ are appealing candidates, as they can be generated with high fidelity using deterministic single-photon emitters, such as quantum dots, as well as probabilistic heralded sources based on spontaneous parametric down-conversion.

Alternative strategies involve manipulating non-Gaussian resource states with Gaussian operations, such as displacement and squeezing~\cite{lvovsky2002synthesis,fiuravsek2005conditional,nielsen2007transforming,miwa2014exploring}, or employing conditional homodyne detection~\cite{ourjoumtsev2007generation,konno2024logical}. Hybrid approaches that combine Gaussian and non-Gaussian resources followed by conditional measurements have also been extensively investigated, including different types of interactions: passive linear optics ~\cite{lvovsky2002quantum,lance2006quantum,bartley2012multiphoton,hu2016multiphoton,birrittella2018photon,eaton2019non,kuts2022realistic}, further equipped with inline squeezing~\cite{winnel2024deterministic}, and active Gaussian coupling such as two-mode squeezing~\cite{zhang2023high,erkilic2025unified,turek2026multiphotonheralding}. 

\emph{Photon catalysis} (PC) constitutes a representative example of such hybrid protocols. In its simplest form, a Fock state is entangled with a Gaussian state through a beam splitter operation, followed by PNRD on one of the output modes. This approach has been explored for the generation of non-Gaussian states \cite{lvovsky2002quantum,bartley2012multiphoton,hu2016multiphoton,birrittella2018photon,eaton2019non,kuts2022realistic} and for the distillation of entanglement \cite{hu2017continuous}. Recently, numerical results on adaptive non-Gaussian state engineering protocols have also been reported in this setting \cite{Crescimanna2025AdaptivePC}. In addition, multimode extensions to the PC protocol have
also been used to approximate cat states \cite{Crescimanna2024SeedGBSwithFock} and quite general results for exact state preparation via multi-mode PC have also been obtained \cite{aralov2025photoncatalysisgeneralmultimode}, similar in spirit to recent work on the possibility of photonic state preparations \cite{TischlerLinearOptics}.

While PC provides a flexible and experimentally accessible route to generate instances of non-Gaussian states, a systematic understanding of the resources required to approximate specific target states within such protocols remains elusive. In particular, it is not clear how to quantify the minimal non-Gaussian complexity needed to achieve a desired state with high fidelity. To address this, we adopt the notion of stellar rank as a measure of non-Gaussianity, capturing the structural complexity of a quantum state within the stellar representation~\cite{Chabaud_2020,chabaud2021CVadvantages,Hanamura2026BeyondStellarRank}. Its relevance for state engineering has also been explored in the context of generating certifiable non-Gaussian states under realistic experimental constraints~\cite{Provaznik2025StellarRank}. Approximate Gaussian conversion protocols have also been addressed using an approximate version of the stellar rank \cite{Hahn_2026}. This framework enables us to formulate state preparation as a resource optimization problem and to leverage PC (a non-Gaussian conversion protocol) to achieve low-complexity approximations of relevant non-Gaussian states. 

In this work, we investigate two-mode PC as a resource-efficient method for the preparation of squeezed coherent state superpositions (squeezed Schrödinger cat states), in both ideal and lossy settings, using stellar rank analysis and Hilbert space truncation in the Fock basis. Such target states are valuable resources for error correction protocols \cite{Schlegel_2022}, for GKP state engineering \cite{eaton2019non,konno2024logical}, and for the generation of GKP cluster states \cite{banic2025exact}.

This work is organized as follows: first, in Sec.\ \ref{sec:catalysis} we introduce the two-mode PC protocol and present numerical results for the optimal approximation of target squeezed cat states. In Sec.\ \ref{sec:stellar_rank}, we review the stellar rank formalism \cite{Chabaud_2020} in the context of optimal quantum state approximation \cite{Chabaud_2021}. The optimal application of pure PC protocols on the generation of squeezed Schrödinger cat states is explored in Sec.\ \ref{sec:cat_approx} using the stellar rank formalism. In Sec.\ \ref{sec:comp_GBS}, we provide a thorough comparison between the PC protocol and the GBS-like scheme presented in 
Ref.\  \cite{su2019pra} with respect to the probability of generating given target states. Next, in Sec.\ \ref{sec:squeezed_fock}, we evaluate the conditions of the PC protocol that enable the generation of squeezed Fock states suitable for error correction \cite{Bashmakova2025squeezedFockQEC}. Finally, we use the numerical methods employed in Sec.\ \ref{sec:catalysis} 
to explore the PC protocol in the presence of loss channels in Sec.\ \ref{sec:losses}.

\section{Photon catalysis}
\label{sec:catalysis}

\begin{figure}
  \centering
  \includegraphics[width=.5\linewidth]{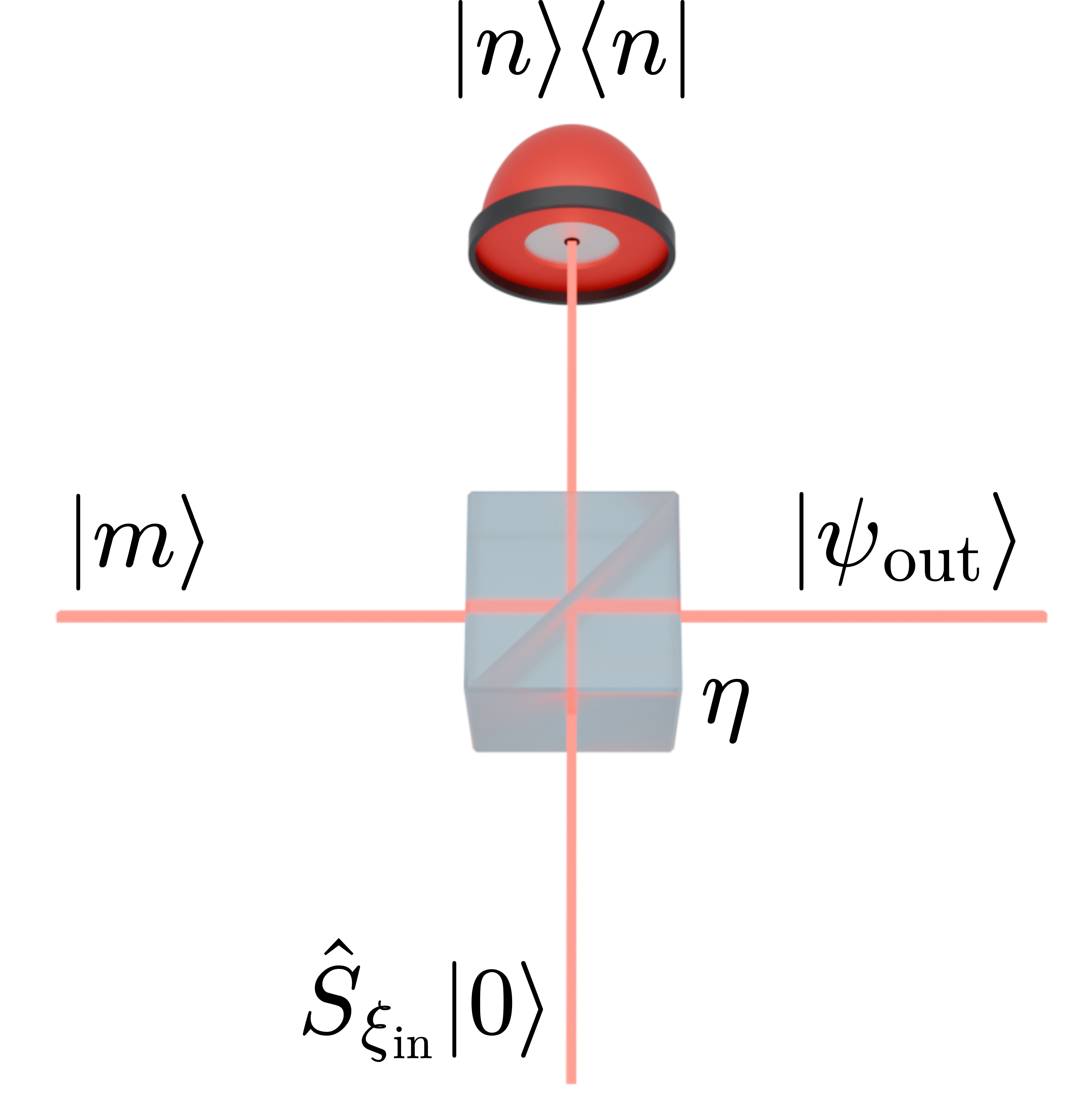}
  \caption{\emph{Photon catalysis} (PC) comprising the interference of a squeezed state vector $\hat{S}_{\xi_\text{in}}\ket{0}$ and a Fock state vector $\ket{m}$ in a beam splitter with transmissivity $\eta$, followed by a PNRD performed at one output port, i.e., a projection onto the Fock state vector $\ket{n}$. The remaining output port yields the state vector $\ket{\psi_\text{out}}$.}
\label{fig:setup}
\end{figure}

The PC protocol of interest consists of a two-mode catalysis between single-mode squeezed vacuum $\hat{S}_{\xi_\text{in}}\ket{0}$ and a Fock state vector $\ket{m}$, followed by the projection onto a Fock state vector $\ket{n}$ in one of the modes \cite{birrittella2018photon}, as represented 
in
Fig.\ \ref{fig:setup}.
The squeezing operator is given by
\begin{equation}
    \hat{S}_{\xi} = \exp\left( \frac{\xi^* \hat{a}^2 - \xi^{} \hat{a}^{\dagger2}}{2}\right),
\end{equation}
where the squeezing magnitude and phase are included in the squeezing parameter $\xi=|\xi|e^{i\theta}$. The transmissivity $\eta$ of the beam splitter is set such that $\eta=0$ and $\eta=1$ refer to $\ket{m}$ and $\hat{S}_{\xi_\text{in}}\ket{0}$ being projected onto $\ket{n}$, respectively. These cases give rise to the output state vector $\ket{\psi_\text{out}}=\hat{S}_{\xi_\text{in}}\ket{0}$ and $\ket{\psi_\text{out}}=\ket{m}$, respectively, provided that the probability of the measurement outcome $n$ does not vanish. Accordingly, the unnormalized output state vector reads as
\begin{align}\label{eq:output_unnorm}
|{\tilde\psi_\text{out}}\rangle = \bra{n} \hat{U}(\eta) \left(\ket{m}\otimes\hat{S}_{\xi_\text{in}}\ket{0}\right) \,,
\end{align}
where the passive unitary $\hat{U}(\eta)$ models the beam splitter operation following the usual transformation relation on the field operators
\begin{align} 
\label{eq:beamsplitter}
    \begin{pmatrix}
    \hat{a}^{\dagger}_{\text{out}_1} \\
    \hat{a}^{\dagger}_{\text{out}_2} 
    \end{pmatrix} =
    \begin{pmatrix}
    \sqrt{\eta} & i\sqrt{1-\eta} \\
    i\sqrt{1-\eta} & \sqrt{\eta}
    \end{pmatrix} \begin{pmatrix}
    \hat{a}^{\dagger}_{\text{in}_1} \\
    \hat{a}^{\dagger}_{\text{in}_2} 
    \end{pmatrix}.
\end{align}
This yields the normalized output state vector as
\begin{align}
\ket{\psi_\text{out}} = \frac{1}{\sqrt{P_\text{PC}}} |{\tilde\psi_\text{out}}\rangle
\end{align}
along with the success probability 
$P_\text{PC}= \langle\tilde\psi_\text{out}| \tilde\psi_\text{out}\rangle$
of detecting $n$ photons in the 
PNRD.
In the Fock basis, the 
squeezed vacuum state vector reads ~\cite{Gerry:book:2005}
\begin{align}
\begin{split}
    \hat{S}_{\xi_\text{in}}\ket{0}=\sum_{q=0}^{\infty}s_{q}\ket{2q},
\end{split}
\end{align}
where
\begin{align}\label{eq:s_m}
	s_{q}=\frac{1}{\sqrt{\cosh{\xi}}}\left(\frac{\tanh{\xi}}{2}\right)^{q}\frac{\sqrt{(2q)!}}{q!}.
\end{align}
Using the beam-splitting transformation in Eq.~\eqref{eq:beamsplitter}, the projected state is 
given by \cite{birrittella2018photon}
\begin{align}\label{eq:Catalyzed_state}
    \ket{\psi_\text{out}}=\sum_{q=0}^{\infty}\sum_{k=0}^{2q}
    Q(2q,m,n,k)\ket{2q+m-n},
\end{align}
with 
\begin{align}\label{eq:Q}
   Q(2q,m,n,k)
    &= \frac{s_{q}}{\sqrt{m!}}
       \begin{pmatrix}2q\\k \end{pmatrix}
       \begin{pmatrix}q\\n-k \end{pmatrix}
       \left(\sqrt{\eta}\right)^{2q-2k+n} \notag \\[4pt]
    & \times\, \left(i\sqrt{1-\eta}\right)^{2k+m-n}
       \sqrt{(2q+m-n)! \, n!}.
\end{align}

Different properties of the presented output state were evaluated, 
including its photon number distribution, generation probability, non-classicality, and non-Gaussian properties \cite{dakna1997generating,dakna1998photon,dakna1998quantum,birrittella2018photon,anaya2021multiphoton}. Such results demonstrate the prospect of heralded protocols in engineering target non-Gaussian states, where high fidelity between PC-like protocols and Schrödinger cat states \cite{kuts2022realistic} and squeezed Schrödinger cat states \cite{eaton2019non} has been demonstrated for particular cases. Here, we determine the optimal catalysis protocol for the generation of squeezed Schrödinger cat target state vectors. They take the form $\hat{S}_{\xi_\text{cat}} \ket{\Ccal_{\alpha_\text{cat}}^\pm}$, where the unsqueezed cat state vector is defined as
\begin{equation}
    \ket{\Ccal_{\alpha_\text{cat}}^\pm} = \frac{1}{\sqrt{C^\pm_{\alpha_\text{cat}}}} \big( \hat{D}_{\alpha_\text{cat}} \pm \hat{D}_{-\alpha_\text{cat}} \big)\ket{0}
\end{equation}
with the cat amplitude $\alpha_\text{cat}$, the parity of the state photon number distribution $\pm$, and the normalization constant $C^\pm_{\alpha_\text{cat}}=2\big(1\pm e^{-2|\alpha_\text{cat}|^2}\big)$.
The displacement operation is defined as
\begin{equation}
    \hat{D}_{\alpha} = \exp \left( \alpha \hat{a}^\dagger - \alpha^*\hat{a} \right),
\end{equation}
with the displacement amplitude and phase contained in $\alpha = |\alpha|e^{i\phi}$. Note that in the definition of a squeezed cat state vector
\begin{align}\label{eq:cat_conv}
    (\hat{S}_{\xi_\text{cat}}\hat{D}_{\alpha_\text{cat}} \pm \hat{S}_{\xi_\text{cat}}\hat{D}_{-\alpha_\text{cat}})\ket{0} \,,
\end{align}
the order of $\hat{S}_{\xi_\text{cat}}$ and $\hat{D}_{\pm\alpha_\text{cat}}$ might deviate from that used in other references. This makes a difference since commuting $\hat{S}_{\xi_\text{cat}}$ and $\hat{D}_{\pm\alpha_\text{cat}}$ modifies the cat amplitude. The convention in Eq.\  \eqref{eq:cat_conv} turns out to facilitate the approximation of squeezed cat states to unsqueezed cat states, particularly in 
Sec.\ \ref{s:Application to cat states}.

For a given protocol with fixed $\xi_\text{in}$, $m$ and $n$, we are interested in determining the optimal splitting parameter $\eta$ that yields the best approximation of the output state to a target squeezed cat state. This task is initially solved by explicitly calculating the fidelity between a list of output states and a list of squeezed cat states. 

In order to reduce the number of states contained in the comparison list, we first note that the parity of the output state is equivalent to the parity of $(m+n)$, as the addition or subtraction of photons will shift the parity of the even number distribution of the squeezed state input. The output state displacement and squeezing phases are also well defined as parallel or orthogonal to the input squeezing phase and in respect to each other, summing up to four different combinations.

By explicitly computing the fidelity between the output state and a list of target states, we can identify the optimal catalysis splitting parameter. The maximum fidelity between a squeezed cat state vector $\hat{S}_{\xi_\text{cat}} \ket{\Ccal_{\alpha_\text{cat}}^\pm}$ and the output state vector $\ket{\psi_\text{out}}$ is chosen as the optimal figure of merit, where $\eta$ and $|\xi_\text{in}|e^{i\theta}$ are kept as free parameters. The calculations were carried out in a truncated Hilbert space of dimension $d=40$. At this dimensionality, a large number of target states can be evaluated in parallel on the GPU of a conventional personal computer. It should be noted, however, that the method does not scale efficiently with either Hilbert space dimension or the number of states included in the comparison. 

\begin{figure}
  \centering
  \includegraphics[width=.75\linewidth]{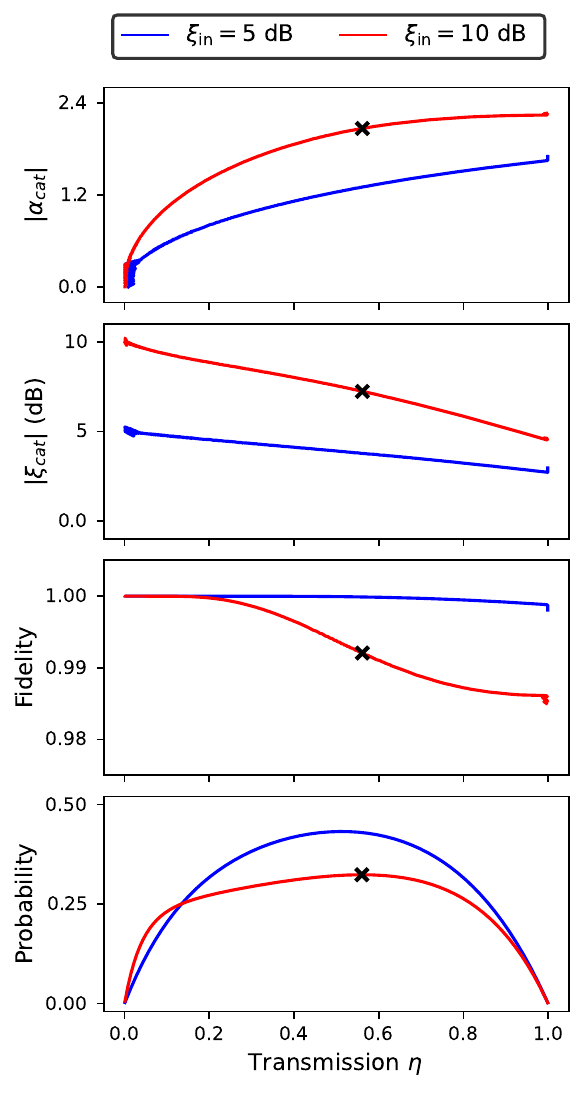}
  \caption{Optimum approximation between the catalysis output state vector $\ket{\psi_\text{out}}$ and target squeezed cat state vectors $\hat{S}_{\xi_\text{cat}} \ket{\Ccal_{\alpha_\text{cat}}^\pm}$ with cat amplitudes in the range of $0<|\alpha_\text{cat}|\leq2.5$. The dependence of the optimal cat amplitude $\alpha_\text{cat}$ and squeezing $\xi_\text{cat}$, the optimal fidelity between $\ket{\psi_\text{out}}$ and $\hat{S}_{\xi_\text{cat}} \ket{\Ccal_{\alpha_\text{cat}}^\pm}$ as well as the success probability on the transmissivity $\eta$ is displayed. The parameters of the considered protocols are $|\xi_\text{in}|=5, 10$ dB, illustrated by different line colors, and $m=2,n=1$.
  The example state indicated with the ``$\mathbf{x}$'' is shown in Fig.\ \ref{fig:HUBn2m1exampleDM}.}
\label{fig:HUBn2m1example}
\end{figure}

\begin{figure}
  \centering
  \includegraphics[width=.85\linewidth]{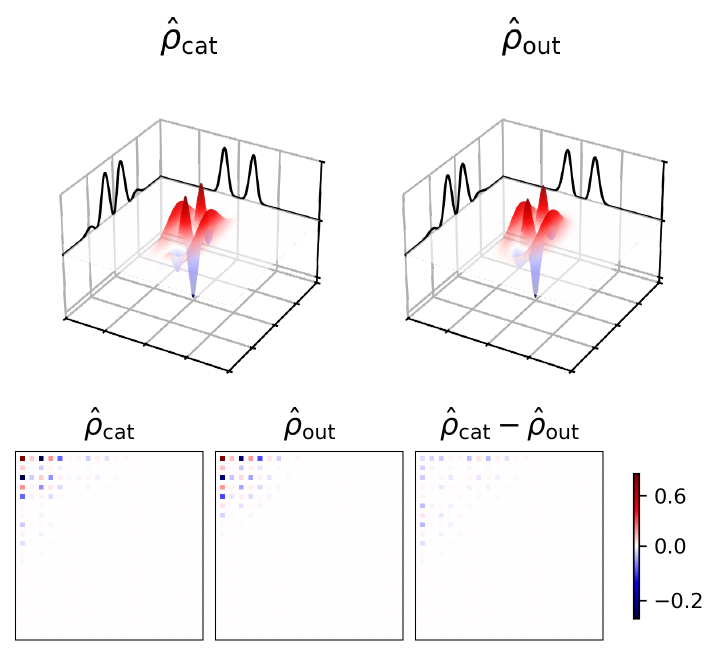}
  \caption{Top: Wigner representation of the target state indicated by the ``$\mathbf{x}$'' in Fig.\  \ref{fig:HUBn2m1example} and the catalysis output that approximates this state with a fidelity of $\mathcal{F}=99.26 \; \%$. Bottom: Density matrix elements of the states represented in the top graphics and the difference between them. The scale is only referent to the matrix elements and uses two different norms in respect to zero for better visualization.
  }
\label{fig:HUBn2m1exampleDM}
\end{figure}

As an illustrative example, we consider $m=2$, $n=1$, and two different input squeezing values: $|\xi_\text{in}|=5,10$ dB. A total of $1000$ output states were compared against $4 \times 1000^2$ squeezed cat states $\hat{S}_{\xi_\text{cat}} \ket{\Ccal_{\alpha_\text{cat}}^\pm}$ with $0<|\alpha_\text{cat}|\leq 2.5$ and  $0\leq|\xi_\text{cat}|\leq12$ dB. The optimal parameters, the maximum fidelity $|\bra{\psi_\text{out}}\hat{S}_{\xi_\text{cat}} \ket{\Ccal_{\alpha_\text{cat}}^\pm}|^2$, and the success probability $P_\text{PC}$ are shown in Fig.\ \ref{fig:HUBn2m1example}. In this example, the optimal displacement and squeezing directions of the output state remained constant and parallel with respect to each other. We also observe that, in the low transmissivity limit $(\eta \to 0)$, multiple parameter combinations yield similarly high fidelities owing to the small amplitude of the output cat state, which gives rise to the ripples observed in the amplitude plot of Fig.\ \ref{fig:HUBn2m1example}. Conversely, larger cat amplitudes are better approximated  in the high transmissivity limit $(\eta \to 1)$. This leads to the abrupt fidelity drop observed at high transmissivities, bounded by the chosen target state set.
Fig.\ \ref{fig:HUBn2m1exampleDM} shows the Wigner function of the target and output states indicated in Fig.\ \ref{fig:HUBn2m1example} by the ``$\mathbf{x}$'', as well as their density matrix elements and the difference between them. 

By adopting a sufficiently large Hilbert space truncation, the proposed method enables a detailed investigation of the generation of squeezed cat states via optical catalysis. However, this approach is computationally demanding, scales poorly, and is susceptible to numerical artifacts arising from finite numerical precision and Hilbert space truncation. To overcome these limitations, we will employ the stellar rank formalism in the approximation of non-Gaussian states. Nevertheless, the truncated-space calculations presented here will be used later in Sec.\ \ref{sec:losses} to analyze the impact of losses in the photon-catalysis protocol.

\section{Stellar rank}
\label{sec:stellar_rank}

Non-Gaussian states are defined in contrast to Gaussian states, which are always described by a Gaussian function on the Wigner or Husimi quasi-probability representations \cite{schleich2015quantum}. For pure states, the presence of zeros in the Husimi distribution (or equivalently negative values in the Wigner representation) is a necessary and sufficient condition for non-Gaussianity. In the single-mode regime, the zeros of the Husimi function form a discrete set that allows a hierarchical categorization of non-Gaussian states based on the stellar rank formalism \cite{Chabaud_2020}. It is shown that such discrimination is equivalent to the number of photon additions or subtractions necessary in order to engineer non-Gaussian states from Gaussian resources. These properties hold for pure and mixed states.

In this section, we will provide a formal introduction to the stellar rank representation followed by a fidelity optimization tool which we have used to characterize the catalysis output state in terms of squeezed cat state approximations.

\subsection{Definition}

In this section, we define the stellar rank as it is usually defined in the literature -- and will turn to manifestly continuous 
definitions later. For pure state vectors $\ket{\psi}$ defined on the Hilbert space of bosonic modes, the stellar rank is formally defined as the number of zeros in its Husimi Q-function, $Q_{\ket{\psi}}(\alpha)=|\braket{\alpha}{\psi}|^2$ \cite{Chabaud_2020}. In this article, we use an equivalent definition that facilitates analytical and numerical calculations. 
A pure state vector $\ket{\psi}$ has stellar rank $N<\infty$ if it can be decomposed as
\begin{align}\label{eq:def_star_N}
    \ket{\psi} = \hat{G}\ket{\phi}
\end{align}
into a Gaussian unitary operator $\hat{G}$ and a so-called \emph{core state} vector $\ket{\phi}$ with Fock support $N$, that is,
\begin{align}\label{eq:def_core_state}
    \ket{\phi}=\sum_{m=0}^N c_m \ket{m}
\end{align}
and $c_N\neq0$. In the case of $N=0$, the state vector $\ket{\psi} = \hat{G}\ket{0}$ is Gaussian. If the state vector cannot be decomposed as in 
Eq.\  \eqref{eq:def_star_N}, that is, if no core state with finite Fock support exists, a stellar rank of $\infty$ is assigned to $\ket{\psi}$. In this sense, $N=\infty$ corresponds to the complement of the Hilbert space comprising all finite-rank states. Even for $N=\infty$, the core state can be defined based on the stellar function \cite{Chabaud_2020}, which is of minor importance for our purpose and thus not detailed here.

In general, the decomposition in Eq.\ \eqref{eq:def_star_N} is not unique since rotation operators $\hat{G}=\hat{R}_\phi$ leave the Fock support invariant and can thus be absorbed into the core state. For displacement and squeezing operators $\hat{G}=\hat{S}_\xi\hat{D}_\alpha$ and finite stellar rank $N<\infty$, however, the uniqueness holds as shown in Ref.\ \cite{Chabaud_2020}. More precisely, if $\ket{\phi}$ and $\ket{\phi'}$ have finite Fock support and fulfill
\begin{align}\label{eq:corestate_unique}
    \hat{D}_\alpha\hat{S}_\xi\ket{\phi} = \hat{D}_{\alpha'}\hat{S}_{\xi'}\ket{\phi'},
\end{align}
then $\alpha={\alpha'}$, $\xi=\xi'$, and $\ket{\phi} = \ket{\phi'}$. This statement will be used in Sec.\ \ref{sec:cat_approx} for squeezing operators. More generally for any state vectors $\ket{\psi}$ and $\ket{\psi'}=\hat{G}\ket{\psi}$, their Fock supports $N$ and $N'$ fulfill $N=N'$ or $N=\infty$ or $N'=\infty$. In particular, a Gaussian unitary operator $\hat{G}$ cannot change the Fock support by a finite amount.

To prove the uniqueness in Eq.\ \eqref{eq:corestate_unique}, we consider $\xi'=\alpha'=0$ without loss of generality, which becomes apparent from applying the inverse of $\hat{D}_{\alpha'}\hat{S}_{\xi'}$ to Eq.\ \eqref{eq:corestate_unique}. Since $\ket{\phi'}$ has finite Fock support $N'$, we know that $\hat{a}^{N'+1}\ket{\phi'}=0$. This yields
\begin{align}
    0&=\hat{a}^{N'+1} \hat{D}_\alpha\hat{S}_\xi\ket{\phi}
    \notag\\&=\hat{D}_\alpha(\hat{a}+\alpha)^{N'+1}\hat{S}_\xi\ket{\phi}
    \notag\\&=\hat{D}_\alpha\hat{S}_\xi (c\hat{a}-s\hat{a}^\dagger+\alpha)^{N'+1}\ket{\phi} \,,
\end{align}
where $c=\cosh{|\xi|}$ and $s=\frac{\xi}{|\xi|}\sinh{|\xi|}$. Hence,
\begin{align}\label{eq:state_vanish}
    (c\hat{a}-s\hat{a}^\dagger+\alpha)^{N'+1}\ket{\phi}=0 \,.
\end{align}
This state vector has Fock support at most $N+N'+1$ since $\ket{\phi}$ has finite Fock support $N$. By expanding the polynomial $(c\hat{a}-s\hat{a}^\dagger+\alpha)^{N'+1}$, the Fock coefficient $N+N'+1$ arises from the term $(-s\hat{a}^\dagger)^{N'+1}\ket{\phi}$. Since it needs to vanish and $(\hat{a}^\dagger)^{N'+1}\ket{\phi}\neq0$, we obtain $s=0$, that is, $\xi=0$. Equation \eqref{eq:state_vanish} thus simplifies to $(\hat{a}+\alpha)^{N'+1}\ket{\phi}=0$, where the state vector has Fock support at most $N$. The Fock coefficient $N$ arises from the term $\alpha^{N'+1}\ket{\phi}$. Since it needs to vanish and $\ket{\phi}\neq0$, we obtain $\alpha=0$, which concludes the proof.

The extension of the stellar rank to mixed states is governed by the convex roof construction, which is nothing but the convex hull: 
For a specific decomposition of the state into a convex combination of pure states, the largest stellar rank within the decomposition is selected. Since the decomposition is not unique, we minimize this value over all decompositions, which defines the stellar rank of a mixed state. 

The stellar hierarchy exhibits limited robustness \cite{Chabaud_2020}, and a more suitable robustness measure is discussed below:
\begin{enumerate}
    \item\label{enum:finite_below} Finite stellar rank states are robust against approximations with states of lower stellar rank.
    
    \item\label{enum:finite_above} 
    Finite stellar rank states can be approximated with states of higher stellar rank arbitrarily well.
    
    \item\label{enum:infinite}
    Infinite stellar rank states can be approximated with states of finite stellar rank arbitrarily well.
\end{enumerate}

To see this for pure states, it suffices to reduce the stellar rank to the Fock support of the respective core states since the decomposition in Eq.\  \eqref{eq:def_star_N} is unique. Claim \ref{enum:finite_below} follows from the largest Fock component $c_N\ket{N}$ with $c_N\neq0$, which cannot be approximated with core states of lower Fock support. For claim \ref{enum:finite_above}, we approximate a core state vector $\ket{\phi}$ of Fock support $N$ with $\ket{\phi}+\veps\ket{N+1}$ and $\veps\to0$. To show claim \ref{enum:infinite}, we truncate the state vector $\ket{\psi}$ to Fock support $N$ and let $N\to\infty$. The statements for mixed states follow from the convex roof construction.

To overcome these limitations in robustness, we will consider the fidelity with states of bounded stellar rank in the subsequent section. This quantity exhibits improved robustness and will be a central figure of merit in the remainder of this article.

\subsection{Optimal approximation}
In quantum photonics, the stellar rank can be interpreted as the minimal number of photon additions and subtractions needed to generate $\ket{\psi}$ from a Gaussian state vector. Thus, the stellar rank formalism is a powerful tool for estimating the required resources for state preparation. Accordingly, the stellar rank of the catalysis protocol output state vector $\ket{\psi_\text{out}}$ is given by $m+n$. However, many desired target states, such as cat and GKP states, have stellar rank $\infty$. 
Hence, we rely on approximating a target state vector $\ket{\psi^*}$ with states of bounded stellar rank determined by the available resources. To this end, we aim to find the optimal state vector $\ket{\psi_N}$ of finite stellar rank $N$ that maximizes the fidelity to the target state vector $\ket{\psi^*}$. Chabaud et al.\  \cite{Chabaud_2021} simplified this task to an optimization over one Gaussian unitary operator $\hat{G}=\hat{S}_\xi\hat{D}_\alpha$ (i.e., two complex variables $\xi,\alpha$). This gives rise to the so-called \emph{stellar fidelity}
\begin{align}\label{eq:stellar_fid}
\Fcal_N(\ket{\psi^*}) &= \sup_{\ket{\psi_N}} |\braket{\psi_N}{\psi^*}|^2
= \sup_{\hat{G}} \Vert \hat\Pi_N \hat{G}\ket{\psi^*} \Vert_2^2 \,,
\end{align}
where $\hat\Pi_N=\sum_{n=0}^N \ket{n}\bra{n}$ is the projection onto Fock support $N$. Note that $\hat{G}$ need not include rotation operators $\hat{R}_\phi$ as $\hat\Pi_N$ is rotationally symmetric. Stellar fidelities have been explored by Hahn et al.\  \cite{Hahn_2026} in conjunction with an approximate version of the stellar rank, which will be discussed in Sec.\ \ref{s:Application to cat states}. The state vector $\ket{\psi_N}$ saturating the stellar fidelity is given by
\begin{align}\label{eq:stellar_state}
\ket{\psi_N} = \hat{G}_\text{opt}^\dagger \ket{\phi_N}
\end{align}
with an optimal Gaussian unitary operator $\hat{G}_\text{opt}$ found in Eq.\  \eqref{eq:stellar_fid} and its core state vector 
\begin{align}
\ket{\phi_N} \propto \hat\Pi_N \hat{G}_\text{opt}\ket{\psi^*} \,.
\end{align}
In line with the notion of stellar fidelities, we call $\ket{\psi_N}$ the \emph{stellar state vector} of $\ket{\psi^*}$. Since $\ket{\phi_N}$ has Fock support of at most $N$ due to the projection $\hat\Pi_N$, the stellar state vector $\ket{\psi_N}$ has at most stellar rank $N$.

For a phase space description of the stellar state vector $\ket{\psi_N} \propto \hat{G}_\text{opt}^\dagger \hat\Pi_N \hat{G}_\text{opt}\ket{\psi^*}$, we interpret the operator $\hat{G}^\dagger \hat\Pi_N \hat{G}$ as Fock projection modified by a Gaussian unitary operator. In phase space, the Wigner function $W_{\hat\Pi_N}$ of $\hat\Pi_N$ has main support on a circle of radius $\sim\sqrt{N}$. Hence, the Wigner function $W_{\ket{\psi^*}}$ of $\ket{\psi^*}$ undergoes a circular projection but with the additional freedom of displacement and squeezing, which yields an elliptical projection. The optimization in Eq.\  \eqref{eq:stellar_fid} thus amounts to finding the largest overlap of $W_{\ket{\psi^*}}$ with an elliptical projection $W_{\hat{G}^\dagger \hat\Pi_N \hat{G}}$. This intuition is also illustrated in Fig.\  \ref{fig:cat_blobs_Wigner}, which is part of the subsequent section, where the optimization in Eq.\  \eqref{eq:stellar_fid} is carried out for cat states.

The stellar fidelity $\Fcal_N(\ket{\psi^*})$ is invariant under Gaussian unitary operators $\hat G'$. This follows from substituting $\hat{G}\mapsto\hat{G}(\hat{G}')^\dagger$ in Eq.\  \eqref{eq:stellar_fid} to give
\begin{align}\label{eq:opt_fid_GU}
    \Fcal_N(\hat{G}'\ket{\psi^*})
    &= \sup_{\hat{G}} \Vert \hat\Pi_N \hat{G}\hat{G}'\ket{\psi^*} \Vert_2^2
    \notag\\&= \sup_{\hat{G}} \Vert \hat\Pi_N \hat{G}\ket{\psi^*} \Vert_2^2
    \notag\\&= \Fcal_N(\ket{\psi^*}) \,.
\end{align}
To derive the optimal state vector $\ket{\psi'_N}$ of $\hat{G}'\ket{\psi^*}$, we apply the same substitution $\hat{G}_\text{opt}\mapsto\hat{G}_\text{opt}(\hat{G}')^\dagger$ to the optimal state vector
\begin{align}\label{eq:opt_state_GU}
    \ket{\psi'_N} &\propto (\hat{G}_\text{opt}(\hat{G}')^\dagger)^\dagger \hat\Pi_N \hat{G}_\text{opt}(\hat{G}')^\dagger \hat{G}'\ket{\psi^*}
    \notag\\&= \hat{G}'\hat{G}_\text{opt}^\dagger \hat\Pi_N \hat{G}_\text{opt}\ket{\psi^*} \,.
\end{align}
We obtain $\ket{\psi'_N}=\hat{G}'\ket{\psi_N}$, where $\ket{\psi_N}$ is the stellar state vector of $\ket{\psi^*}$. Hence, modifications of the target state vector $\ket{\psi^*}$ by Gaussian unitary operators lead to the same modifications of the stellar state vector $\ket{\psi_N}$. This perfectly matches the previously described intuition in phase space. If $W_{\ket{\psi^*}}$ is modified by a Gaussian unitary operator, then $W_{\hat{G}^\dagger \hat\Pi_N \hat{G}}$ also needs to be modified in order to preserve their overlap. This is equivalent to transforming the phase space coordinates, which equally transforms the stellar state.

As aforementioned, the stellar fidelity $\Fcal_N(\ket{\psi^*})$ in Eq.\  \eqref{eq:stellar_fid} exhibits improved robustness compared to the stellar rank itself. Small variations of $\ket{\psi^*}$ solely lead to small variations of $\Fcal_N(\ket{\psi^*})$, which follows from continuity of the norm $\Vert.\Vert_2$ and the supremum function. Importantly, this holds irrespective of the stellar rank of $\ket{\psi^*}$ and whether the variation includes states with lower or higher stellar rank.

\subsection{Application to cat states}\label{s:Application to cat states}
In this section, we use the framework of the preceding section to approximate squeezed cat states with states of finite stellar rank. Fortunately, this approximation is possible for cat states since infinite stellar rank states are not robust (cf. claim \ref{enum:infinite}). In other words, $\Fcal_N(\ket{\psi^*})\to1$ for $N\to\infty$. The main results of this section have already been discovered by Chabaud~\cite{chabaud2021CVadvantages}. In view of the comparison to PC, we elaborate on them in more detail. Finite stellar rank approximations of cat states consisting of more than two coherent components have been studied by Wang et al.~\cite{wang2026bosonicQEC}. A more general approach to relate finite stellar rank to non-Gaussian features is provided by Provaznik et al.~\cite{provaznik2026witnessesnongaussian}.

To begin with, we set the target state to $\ket{\psi^*}=\hat{S}_{\xi_\text{cat}}\ket{\Ccal_{\alpha_\text{cat}}^\pm}$ in Eq.\  \eqref{eq:stellar_fid} with the cat state vector 
\begin{equation}
\ket{\Ccal_{\alpha_\text{cat}}^\pm}\propto (\hat{D}_{\alpha_\text{cat}}\pm\hat{D}_{-\alpha_\text{cat}})\ket{0}.
\end{equation}
To compute the stellar fidelity $\Fcal_N(\ket{\psi^*})$ and the stellar state vector $\ket{\psi_N}$, we first simplify this task to unsqueezed cat state vectors $\ket{\Ccal_{\alpha_\text{cat}}^\pm}$. As shown in Eq.\  \eqref{eq:opt_fid_GU}, the stellar fidelity is invariant under $\hat{S}_{\xi_\text{cat}}$ as
\begin{align}
    \Fcal_N(\hat{S}_{\xi_\text{cat}}\ket{\Ccal_{\alpha_\text{cat}}^\pm}) = \Fcal_N(\ket{\Ccal_{\alpha_\text{cat}}^\pm})
\end{align}
and thus independent of the cat squeezing $\xi_\text{cat}$. In the following, we abbreviate the stellar fidelity of cat states as $\Fcal_N= \Fcal_N(\ket{\Ccal_{\alpha_\text{cat}}^\pm})$. As we show in Appendix \ref{sec:appendix_opt_fid}, $\Fcal_N$ can be computed efficiently without truncation in the Fock basis, in contrast to the numerical approach described in Sec.\  \ref{sec:catalysis}. According to Eq.\  \eqref{eq:opt_state_GU}, $\hat{S}_{\xi_\text{cat}}\ket{\psi_N}$ is the stellar state vector of a squeezed cat state vector $\hat{S}_{\xi_\text{cat}}\ket{\Ccal_{\alpha_\text{cat}}^\pm}$, where
\begin{align}\label{eq:stellar_state_cat}
    \ket{\psi_N}\propto (\hat{G}_\text{opt})^\dagger \hat\Pi_N \hat{G}_\text{opt}\ket{\Ccal_{\alpha_\text{cat}}^\pm}
\end{align}
is the stellar state vector of the unsqueezed cat state vector $\ket{\Ccal_{\alpha_\text{cat}}^\pm}$. 

We point out the case $N=0$, where the stellar fidelity $\Fcal_0$ is called the \emph{Gaussian fidelity} and represents the maximum achievable fidelity with Gaussian states. This has been investigated by Lami et al.\  \cite{Lami2021resourceinfdim} and Hahn et al.\  \cite{Hahn2025classicalsimulation}, also in conjunction with the Gaussian extent. The Gaussian fidelity serves as a threshold for the output state vector $\ket{\psi_\text{out}}$ to fundamentally justify PC in contrast to simple Gaussian states. Namely, the fidelity of $\ket{\psi_\text{out}}$ and a target squeezed cat state ought to exceed $\Fcal_0$ since otherwise, the same result can equally be achieved with Gaussian states whose generation is deterministic and much less expensive. Moreover, we expect that cat codes only provide quantum advantage if $\Fcal_0$ is exceeded. The Gaussian fidelity will become important in Sec.\  \ref{sec:losses}, where we investigate the influence of loss.

\begin{figure}
  \centering
  \includegraphics[width=.75\linewidth]{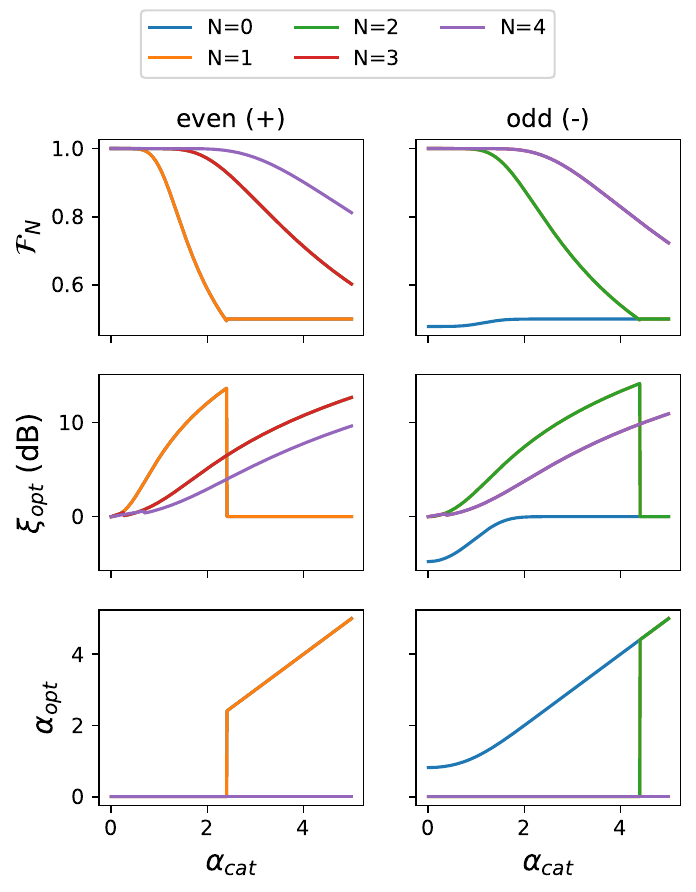}
  \caption{
  Dependence of the stellar fidelity $\Fcal_N$, i.e., the maximum achievable fidelity between the cat state vector $\ket{\Ccal_{\alpha_\text{cat}}^\pm}$ and states of at most stellar rank $N$, as well as the squeezing and displacement parameters $\xi_\text{opt},\alpha_\text{opt}$ of the optimal Gaussian unitary operator $\hat{G}_\text{opt} = \hat{S}_{\xi_\text{opt}}\hat{D}_{\alpha_\text{opt}}$, on the cat amplitude $\alpha_\text{cat}$. This is displayed for even (+) and odd (-) cat states as well as for stellar rank $N\leq4$. In all cases, except for odd cat states and $N=0$, a phase transition occurs at the threshold $z^\pm_N$, listed in Tab.\ \ref{tab:z_N}. At $\alpha_\text{cat}=z^\pm_N$, the stellar fidelity reaches $\Fcal_N=1/2$ and the optimal Gaussian unitary operator changes from $\hat{G}_\text{opt} = \hat{S}_{\xi_\text{opt}}$ to $\hat{G}_\text{opt} = \hat{D}_{\alpha_\text{cat}}$. The Gaussian fidelity is given by $\Fcal_0$ for even and odd cat states.}
  \label{fig:opt_state}
\end{figure}

Next, we compute $\Fcal_N$ and the corresponding optimal Gaussian unitary operator $\hat{G}_\text{opt} = \hat{S}_{\xi_\text{opt}}\hat{D}_{\alpha_\text{opt}}$. In Fig.~\ref{fig:opt_state}, we present the dependence of $\Fcal_N,\xi_\text{opt},\alpha_\text{opt}$ on the cat amplitude $\alpha_\text{cat}$ for even ($\ket{\Ccal_{\alpha_\text{cat}}^+}$) and odd ($\ket{\Ccal_{\alpha_\text{cat}}^-}$) cat states as well as for stellar rank $N\leq4$, including the Gaussian fidelity $\Fcal_0$. The cat amplitude is set to be real-valued $\alpha_\text{cat}>0$ while complex cat amplitudes simply follow from rotation in phase space. 
Moreover, we infer the highest achievable cat amplitude for lower bounded stellar fidelity $\Fcal_N\geq0.99,0.999,0.9999$ and fixed stellar rank $N$, which is presented in Tab.\ \ref{tab:fid_cut}. This corresponds to the approximate stellar rank, which was defined by Hahn et al.\  \cite{Hahn_2026} and initially stems from the $\varepsilon$-smoothed non-Gaussianity of formation, introduced by Chabaud et al.\  \cite{Chabaud_2020}. The $\varepsilon$-approximate stellar rank indicates the minimal stellar rank $N$ such that the stellar fidelity fulfills $\Fcal_N\geq1-\varepsilon$. This notion is suitable to assess the stellar rank in view of error correction. For example, a cat code with $\alpha=0.55$ considered by Schlegel et al.\  \cite{Schlegel_2022}, which is $\alpha_\text{cat}\approx 2.46$ in our notation, requires stellar rank $N\geq 4$ to guarantee fidelity $\Fcal_N\geq 0.99$. In other words, the code has $0.01$-approximate stellar rank $4$. For a broader discussion of error correction features of finite stellar rank approximations, we refer to~\cite{wang2026bosonicQEC}.

\begin{table}
\centering
{\renewcommand{\arraystretch}{1.2}


$\Fcal_N\geq0.99:\;$
\begin{tabular}{ccccccccccccc}
\hline\hline
$N$ && 0 && 1 && 2 && 3 && 4 && 5
\\ \hline
$|\alpha_\text{cat}|\leq$ && 0.74 && 1.20 && 1.68 && 2.13 && 2.57 && 2.98
\\\hline\hline
\end{tabular}

$\ph$\\$\ph$\\$\Fcal_N\geq 0.999:\;$
\begin{tabular}{ccccccccccccc}
\hline\hline
$N$ && 0 && 1 && 2 && 3 && 4 && 5
\\ \hline
$|\alpha_\text{cat}|\leq$ && 0.54 && 0.86 && 1.26 && 1.61 && 1.95 && 2.29
\\\hline\hline
\end{tabular}

$\ph$\\$\ph$\\$\Fcal_N\geq0.9999:\;$
\begin{tabular}{ccccccccccccc}
\hline\hline
$N$ && 0 && 1 && 2 && 3 && 4 && 5
\\ \hline
$|\alpha_\text{cat}|\leq$ && 0.40 && 0.64 && 1.00 && 1.28 && 1.58 && 1.86
\\\hline\hline
\end{tabular}
}
\caption{Dependence of the upper bound of the cat amplitude $|\alpha_\text{cat}|$ for lower bounded stellar fidelity $\Fcal_N$, i.e., the maximum achievable fidelity between the cat state vector $\ket{\Ccal_{\alpha_\text{cat}}^\pm}$ and states of at most stellar rank $N$, on the stellar rank $N$. The values stem from Fig.\  \ref{fig:opt_state}.}
\label{tab:fid_cut}
\end{table}

\begin{table}
\centering
{\renewcommand{\arraystretch}{1.2}
\begin{tabular}{ccccccccccc}
\hline\hline
$N$ && 0 && 1 && 2 && 3 && 4
\\ \hline
$z^+_N$ && 2.373 && 2.373 && 6.248 && 6.248 && 9.670
\\ \hline
$z^-_N$ && - && 4.373 && 4.373 && 8.007 && 8.007
\\\hline\hline
\end{tabular}
}
\caption{Dependence of the threshold $z^\pm_N$ pertaining to the phase transition in Fig.\  \ref{fig:opt_state} on the stellar rank $N$ and the cat state parity $\pm$. At $|\alpha_\text{cat}|=z^\pm_N$, the 
stellar fidelity reaches $\Fcal_N=1/2$ and the optimal Gaussian unitary operator changes from $\hat{G}_\text{opt} = \hat{S}_{\xi_\text{opt}}$ to $\hat{G}_\text{opt} = \hat{D}_{\alpha_\text{cat}}$. For odd cat states and $N=0$, no phase transition occurs.}
\label{tab:z_N}
\end{table}

To derive these results, we start with even cat state vectors $\ket{\Ccal_{\alpha_\text{cat}}^+}$ as targets. Numerical investigations reveal two essential regimes, which occur depending on the cat amplitude $\alpha_\text{cat}$: (A) For small $\alpha_\text{cat}$, $\alpha_\text{opt}=0$ with $\xi_\text{opt}>0$. (B) For large $\alpha_\text{cat}$, $\alpha_\text{opt}=\alpha_\text{cat}$ with $\xi_\text{opt}=0$. In these regimes, the optimal Gaussian unitary operator approximately becomes (A) a squeezing $\hat{G}_\text{opt} = \hat{S}_{\xi_\text{opt}}$ and (B) a displacement $\hat{G}_\text{opt} = \hat{D}_{\alpha_\text{cat}}$. The optimal solution changes between these regimes if $|\alpha_\text{cat}|$ surpasses a threshold $z^\pm_N$, which is listed in Tab.\ \ref{tab:z_N} and depends on the stellar rank $N$ as well as on the cat state parity $\pm$.

\begin{figure}
  \centering
  \includegraphics[width=.8\linewidth]{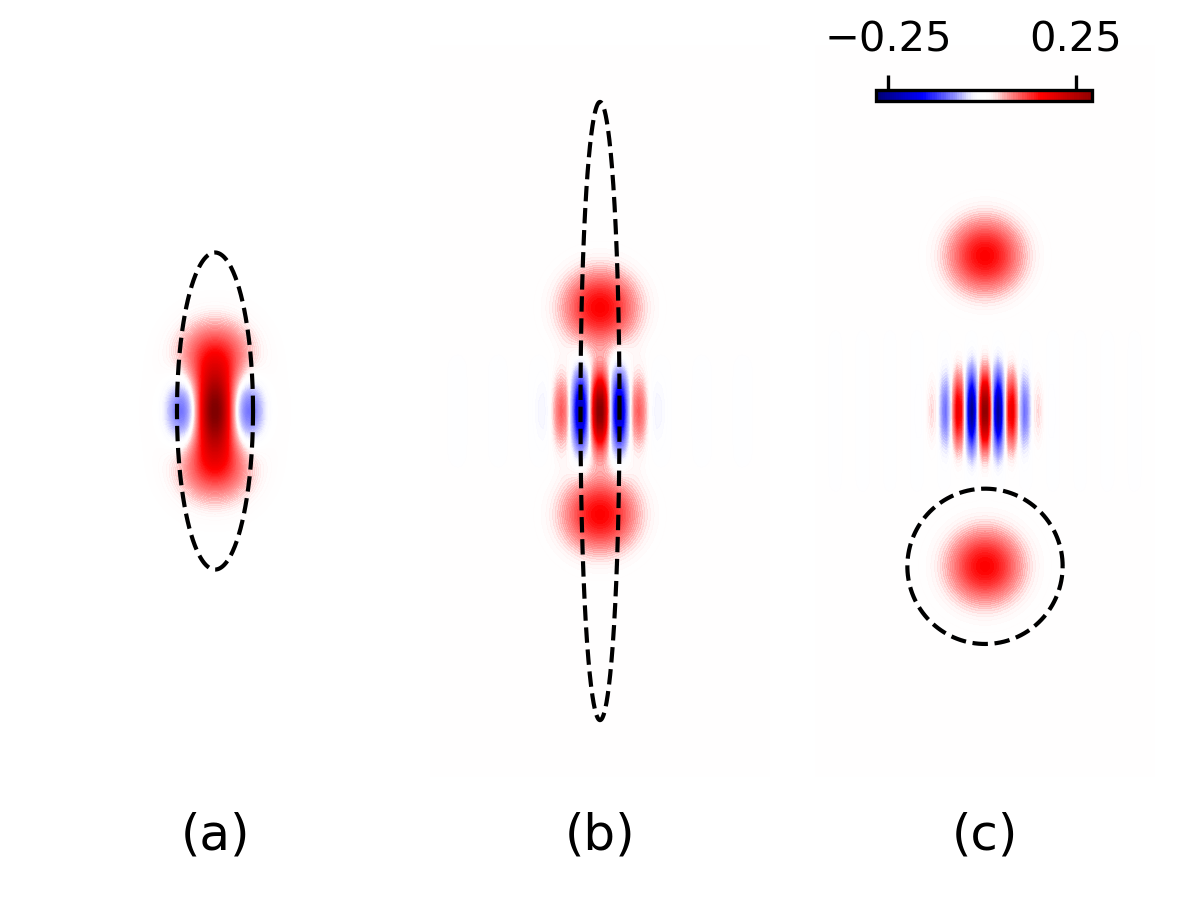}
  \caption{Representations of Wigner functions of even cat states with different amplitudes (color maps) are compared with the optimal Gaussian $\hat{G}_\text{opt} = \hat{S}_{\xi_\text{opt}}\hat{D}_{\alpha_\text{opt}}$, depicted as the $3\sigma$ contour of the quadrature distribution of the state $\hat{G}_\text{opt}\ket{0}$ (dashed lines). In the example, the stellar rank bound is $N=0$, where the cat state amplitudes are (a) $|\alpha_\text{cat}|=1<z^+_0$, with $|\alpha_\text{opt}|=0$ and $|\xi_\text{opt}|=6.2$ dB; (b) $|\alpha_\text{cat}|=2<z^+_0$, with $|\alpha_\text{opt}|=0$ and $|\xi_\text{opt}|=12$ dB and (c) $|\alpha_\text{cat}|=3>z^+_0$, with $|\alpha_\text{opt}|=3$ and $|\xi_\text{opt}|=0$. The threshold is given by $z^+_0=2.4$.}
\label{fig:cat_blobs_Wigner}
\end{figure}

For an illustrative explanation, we again portray the situation in phase space. The following description is also illustrated in Fig.\  \ref{fig:cat_blobs_Wigner} for $N=0$, i.e., 
optimal approximation with Gaussian states. The Wigner function $W_{\ket{\Ccal_{\alpha_\text{cat}}^+}}$ of an even cat state comprises two Gaussian functions located at $\pm\alpha_\text{cat}$, which pertain to the coherent components $\ket{\pm\alpha_\text{cat}}$ of the cat state, and an interference pattern in between. In the limit of vanishing cat amplitude $\alpha_\text{cat}\to0$, the coherent components merge and form one Gaussian function culminating in the vacuum state. We intend to maximize the overlap of $W_{\ket{\Ccal_{\alpha_\text{cat}}^\pm}}$ with the elliptical projection $W_{\hat{G}^\dagger \hat\Pi_N \hat{G}}$, which is free to be displaced and squeezed. In regime (A) of small $\alpha_\text{cat}$, the coherent components are sufficiently close, such that it is optimal for $W_{\hat{G}^\dagger \hat\Pi_N \hat{G}}$ to cover both coherent components simultaneously. Therefore, $W_{ \hat\Pi_N}$ merely needs to be squeezed and not displaced, hence $\hat{G}_\text{opt} = \hat{S}_{\xi_\text{opt}}$. In particular, the squeezing is directed towards the coherent components, which means $\xi_\text{opt}>0$ for $\alpha_\text{cat}\in\R$ in our notation.

For increasing cat amplitude $\alpha_\text{cat}$, however, the spacing between the coherent components renders the simultaneous coverage more and more impractical, which causes $\Fcal_N$ to decrease. At the threshold $z^+_N$, the optimal solution flips to covering only one coherent component, that is, regime (A) changes to regime (B). In regime (B), the coherent components are sufficiently separated from each other and from the interference pattern, such that one coherent component can be treated as a single coherent state with half amplitude. Covering a coherent state in turn only requires displacement, hence $\hat{G}_\text{opt} = \hat{D}_{\alpha_\text{cat}}$. In particular, the optimization task of approximating a coherent state can be solved optimally just by using Gaussian states. As a consequence for regime (B), the Gaussian fidelity $\Fcal_0$ cannot be exceeded. Since only half of the cat state is covered, we obtain $\Fcal_N=1/2$. The phase transition between (A) and (B) at the threshold $|\alpha_\text{cat}|=z^+_N$ thus occurs when $\Fcal_N$ would have dropped below $1/2$.

The range of the elliptical projection $W_{\hat{G}^\dagger \hat\Pi_N \hat{G}}$ increases with $N$ such that in regime (A), the simultaneous coverage of both coherent components remains optimal even for higher cat amplitudes $\alpha_\text{cat}$. In other words, $\Fcal_N$ increases with $N$ for fixed cat amplitude $\alpha_\text{cat}$ and so does the threshold $z^+_N$. Above this threshold in regime (B), the cat amplitude remains too large to cover more than one coherent component, which prevents higher stellar rank from outperforming approximations with Gaussian states. Moreover, we observe in Fig.\  \ref{fig:opt_state} that for even cat states, $\Fcal_N$ appears to coincide for $N=0,1$ and $N=2,3$. Indeed, the maximum deviation between coinciding curves (of both even and odd cat states) amounts to $8\cdot10^{-11}$, $10^{-5}$ dB, and $7\cdot10^{-5}$ for $\Fcal_N$, $\xi_\text{opt}$, and $\alpha_\text{opt}$, respectively, at most. In regime (A), this becomes plausible from the parity of even cat states. In Eq.\  \eqref{eq:stellar_fid}, squeezing $\hat{G}_\text{opt} = \hat{S}_{\xi_\text{opt}}$ preserves the parity of the target cat state such that $\hat{\Pi}_N$ is applied to a state with even Fock occupation. Adding a projection onto an odd Fock state vector $\ket{N+1}\bra{N+1}$ to $\hat{\Pi}_N$ thus leaves $\Fcal_N$ invariant. In regime (B), the optimization task reduces to approximating a coherent state, which is identical for all stellar rank $N$. The aforementioned deviations simply originate from the numerical optimization, where $\hat{G}_\text{opt} = \hat{S}_{\xi_\text{opt}}$ and $\hat{G}_\text{opt} = \hat{D}_{\alpha_\text{cat}}$ do not hold exactly, e.g., due to slight instabilities or local maxima.

Next, we discuss the odd cat state vectors $\ket{\Ccal_{\alpha_\text{cat}}^-}$ as target state vectors. Many observations that were elucidated for even cat states equally hold for odd cat states. In Fig.\  \ref{fig:opt_state}, we observe the significant ``phase transition'' between the regimes (A) and (B) at the threshold $z^-_N$ for $N\geq1$ along with the optimal Gaussian unitary operator changing from $\hat{G}_\text{opt} = \hat{S}_{\xi_\text{opt}}$ to $\hat{G}_\text{opt} = \hat{D}_{\alpha_\text{cat}}$. However, there are two key differences from even cat states: first, the parity is odd due to their odd Fock occupation and the curves for $N=1,2$ and $N=3,4$ thus coincide, which can be explained as for even cat states. Secondly, the limit for $\alpha_\text{cat}\to0$ reveals the 1-Fock state vector $\ket{\Ccal_{\alpha_\text{cat}}^-}\to\ket{1}$ instead of the vacuum state. States of stellar rank $N\geq1$ perfectly cover $\ket{1}$, such that $\Fcal_N\to1$ for $\alpha_\text{cat}\to0$ as for even cat states. In the case of $N=0$, however, Gaussian states attempt to approximate $\ket{1}$ in the limit of $\alpha_\text{cat}\to0$. As computed by Chabaud et al., the maximum fidelity $\Fcal_0$ amounts to $3\sqrt{3}/(4e)\approx 0.478$ \cite{Chabaud_2020}. In phase space, the optimal Gaussian state is a squeezed, coherent state covering part of the external ring of the Fock state's Wigner function $W_{\ket{1}}$, hence $\alpha_\text{opt}>0$ and $\xi_\text{opt}<0$. Surprisingly, $\Fcal_0$ does not exceed $0.5$, which is obtained in the limit of large cat amplitude $\alpha_\text{cat}$, where the coherent components are sufficiently separated. This is all the more crucial as $\Fcal_0$ represents the Gaussian fidelity for odd cat states. In contrast to any other case, no phase transition occurs but a smooth increase of $\Fcal_0$ with $\alpha_\text{cat}$ from $0.478$ to $0.5$. Similarly, $\alpha_\text{opt}$ and $\xi_\text{opt}$ smoothly converge to a linear function of $\alpha_\text{cat}$ and to $0$, respectively.

\section{Cat states approximation using PC}
\label{sec:cat_approx}

In this section, we compare the output state $\ket{\psi_\text{out}}$ of the PC protocol with the stellar state vector $\ket{\psi_N}$ of cat states, which is defined in Eq.\ \eqref{eq:stellar_state_cat}. For a broad range of parameters ($\xi_\text{in}, \; m, \;n$ and $\eta$ ), the output state coincides with the optimal stellar rank approximation of targeted states with different cat amplitudes. The 
main goal of this section is to find parameters $\eta$ and $\xi_\text{in}$ such that $\ket{\psi_\text{out}}$ is optimal for the generation of some cat state vector, 
i.e.,
\begin{align}\label{eq:opt_exp_output}
\ket{\psi_\text{out}} = \hat{S}_{\xi_\text{cat}}\ket{\psi_N} \,,
\end{align}
where $\ket{\psi_N}$ is the stellar state vector of some unsqueezed cat state vector $\ket{\Ccal_{\alpha_\text{cat}}^\pm}$. Moreover, we require that the Gaussian fidelity $\Fcal_0$ is exceeded to guarantee an actual non-Gaussian advantage. As discussed in Sec.\  \ref{s:Application to cat states}, this is only fulfilled in regime (A). To avoid regime (B), we will thus discard solutions with $\Fcal_N<1/2$, i.e., $|\alpha_\text{cat}|>z^\pm_N$ with the threshold $z^\pm_N$ listed in 
Tab.\ \ref{tab:z_N}.

For simplicity, we consider $\xi_\text{in}\in\R$ in the following since the remaining state vectors $\ket{m},\ket{n}$ of the PC setup are rotationally invariant. Note that real squeezing parameters correspond to squeezing along the position and the momentum axes in phase space. To address Eq.\  \eqref{eq:opt_exp_output}, we begin by writing the output state as
\begin{align}\label{eq:output_dec}
\ket{\psi_\text{out}} = \hat{S}_{\xi_\text{out}} \ket{\phi_\text{out}}
\end{align}
in terms of the decomposition in Eq.\  \eqref{eq:def_star_N}, where the core state vector $\ket{\phi_\text{out}}$ has a stellar rank $N=m+n$. Since the input squeezing is real, so is the output squeezing $\xi_\text{out}$ and given by
\begin{align}\label{eq:xi_out}
\tanh\xi_\text{out}=(1-\eta)\tanh\xi_\text{in}\,,
\end{align}
which is derived in Appendix \ref{sec:appendix_Fock}. On the other hand, the stellar state vector of a cat state is given by
\begin{align}
    \ket{\psi_N} = \hat{S}_{\xi_\text{opt}}^\dagger \ket{\phi_N}
\end{align}
with its core 
state vector $\ket{\phi_N}$ since in regime (A), the optimal Gaussian unitary operator is a squeezing $\hat{G}_\text{opt} = \hat{S}_{\xi_\text{opt}}$. Inserting these relations into 
Eq.\  \eqref{eq:opt_exp_output} yields
\begin{align}
\hat{S}^{\phd}_{\xi_\text{out}} \ket{\phi_\text{out}} = \hat{S}^{\phd}_{\xi_\text{cat}}\hat{S}_{\xi_\text{opt}}^\dagger \ket{\phi_N} \,.
\end{align}
Since the decomposition into a squeezing operator and the core state is unique, discussed in Eq.\ \eqref{eq:corestate_unique}, we separate this problem to
\begin{align}\label{eq:opt_exp_output_decomp}
\hat{S}_{\xi_\text{out}}^{\phd}=\hat{S}_{\xi_\text{cat}}^{\phd} \hat{S}_{\xi_\text{opt}}^\dagger \,,\quad
\ket{\phi_\text{out}} = \ket{\phi_N} \,.
\end{align}
Next, we justify the assumption that the squeezing parameters $\xi_\text{out},\xi_\text{cat},\xi_\text{opt}$ are real. To this end, we first recall that $\xi_\text{out}\in\R$ based on Eq.\  \eqref{eq:xi_out}. In particular, the squeezed input state $\hat{S}_{\xi_\text{in}}\ket{0}$ is invariant under complex conjugation in phase space. Since the remaining state vectors $\ket{m},\ket{n}$ of the PC setup also feature this invariance, so does the output state. In turn, this implies that its core state vector $\ket{\phi_\text{out}}$ has real Fock coefficients $\braket{k}{\phi_\text{out}}\in\R$ for all $k\leq N$. Owing to Eq.\  \eqref{eq:opt_exp_output_decomp}, the same holds for the Fock coefficients of the optimal core state $\braket{k}{\phi_N}\in\R$. This is only possible for real and purely imaginary cat amplitude $\alpha_\text{cat}$, where the coherent components are located along the position and momentum axis in phase space, respectively. Since the optimal squeezing $\hat{S}_{\xi_\text{opt}}$ points towards the coherent components, we conclude that $\xi_\text{opt}\in\R$. This serves to simplify $\hat{S}_{\xi_\text{out}} \hat{S}_{\xi_\text{opt}} = \hat{S}_{\xi_\text{out}+\xi_\text{opt}}^{\phd}$ in Eq.\  \eqref{eq:opt_exp_output_decomp}, which yields
\begin{align}\label{eq:xi_cat}
\xi_\text{cat}=\xi_\text{out}+\xi_\text{opt} \,.
\end{align}
Note that the implication from Eq.\  \eqref{eq:opt_exp_output_decomp} to \eqref{eq:xi_cat} does not hold for arbitrary complex squeezing parameters. It remains to satisfy $\ket{\phi_\text{out}} = \ket{\phi_N}$. Cat state vectors $\ket{\Ccal_{\alpha_\text{cat}}^+}$ and $\ket{\Ccal_{\alpha_\text{cat}}^-}$ have purely even and odd Fock occupation, respectively, and so does $\ket{\phi_N}\propto\hat\Pi_N\hat{S}_{\xi_\text{opt}}\ket{\Ccal_{\alpha_\text{cat}}^\pm}$ since $\Pi_N$ and $\hat{S}_{\xi_\text{opt}}$ preserve the parity. Conversely, $\ket{\phi_\text{out}}$ has purely even and odd Fock occupation if $N$ is even and odd, respectively, due to the parity symmetry of the states occurring in PC. Hence, it suffices to check whether $\lfloor N/2\rfloor+1$ real parameters coincide, namely
\begin{align}\label{eq:coef_coinc}
\braket{k}{\phi_\text{out}} = \braket{k}{\phi_N}
\end{align}
for $k\in\{0,2,\ldots,N\}$ or $k\in\{1,3,\ldots,N\}$, irrespective of a global sign. Once again, numerical computations of Eq.\  \eqref{eq:coef_coinc} are carried out efficiently without Fock truncation. Details for the left-hand and the right-hand sides are provided in Appendices \ref{sec:appendix_opt_fid} and \ref{sec:appendix_Fock}, respectively.

\begin{figure*}
  \centering
  \includegraphics[width=0.75\linewidth]{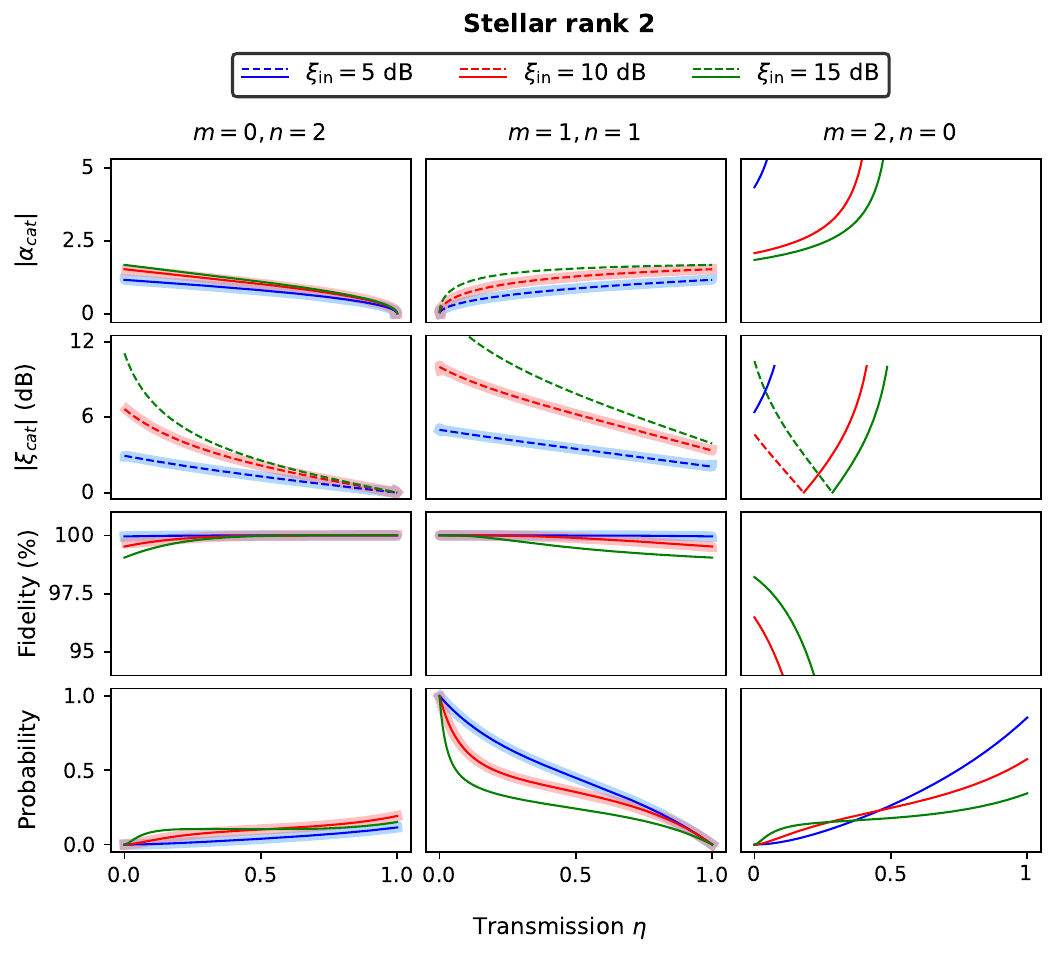}
\caption{Dependence of the cat amplitude
$\alpha_\text{cat}$ and cat squeezing $\xi_\text{cat}$, which satisfy that the output state vector $\ket{\psi_\text{out}}$ is closest to the even cat state $\hat{S}_{\xi_\text{cat}}\ket{\Ccal_{\alpha_\text{cat}}^+}$ among the states of stellar rank $N=2$, on $\eta$ and $\xi_\text{in}$. Respective stellar fidelities and success probabilities are also displayed. Different columns refer to different combinations of $m$ and $n$ such that $m+n=2$. If the output state is not closest to any cat state or $|\alpha_\text{cat}|>z^\pm_N$, the respective values are not displayed.
Different line styles label the direction in phase space: solid/dashed refers to real/imaginary cat amplitude $\alpha_\text{cat}$ in the first row and positive/negative cat squeezing $\xi_\text{cat}$ in the second row. The fidelity and the success probability in the third and forth row, respectively, are plotted with a solid linestyle since this direction in phase space is irrelevant.
Shaded background lines show numerical results from the method described in Sec.\ \ref{sec:catalysis}.}
\label{fig:stellar2_theo}
\end{figure*}

\begin{figure*}
  \centering
  \includegraphics[width=.95\linewidth]{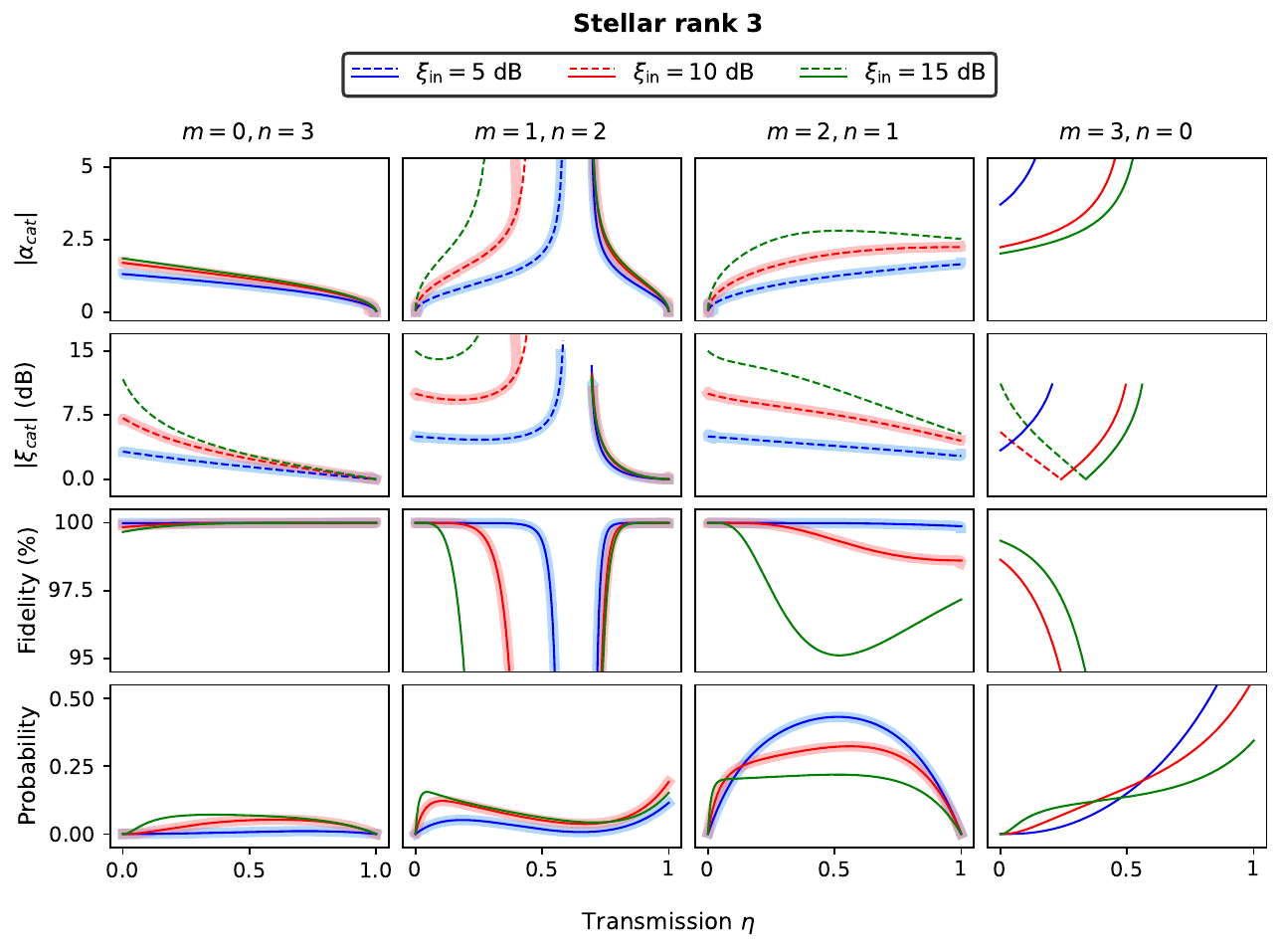}
  \caption{Similar to Fig.\  \ref{fig:stellar2_theo} for output states of stellar rank $N=3$. Different columns refer to different combinations of $m$ and $n$ such that $m+n=3$.}
\label{fig:stellar3_theo}
\end{figure*}

We next discuss several cases for $N$. The cases $N=0$ and $N=1$ lead to trivial core state vectors $\ket{\phi_0}=\ket{0}$ and $\ket{\phi_1}=\ket{1}$, respectively, such that Eq.\  \eqref{eq:coef_coinc} is trivially fulfilled for all $\eta$ and $\xi_\text{in}$. We thus focus on $N=2$ and $N=3$ as the simplest non-trivial cases, where two real parameters in Eq.\  \eqref{eq:coef_coinc} need to coincide. Figures \ref{fig:stellar2_theo} and \ref{fig:stellar3_theo} display the range of $\eta$ and $\xi_\text{in}$, where this is fulfilled for a cat state vector $\ket{\Ccal_{\alpha_\text{cat}}^\pm}$, as well as the corresponding cat amplitude $\alpha_\text{cat}$, cat squeezing $\xi_\text{cat}$, stellar fidelity $\Fcal_N$, and success probability $P_\text{PC}$. In particular, we solely display solutions fulfilling $|\alpha_\text{cat}|<z^\pm_N$, i.e., $\Fcal_N>1/2$. To relate the directions of cat amplitude and cat squeezing, the relevant cases $\alpha_\text{cat}\in\R,\alpha_\text{cat}\in i\R$ and $\xi_\text{cat}\geq 0,\xi_\text{cat}\leq 0$ are accordingly illustrated. In addition, we display the results obtained with the method described in Sec.\ \ref{sec:catalysis} as a background shadow of the curves obtained with the stellar rank optimization method. Despite the results being in good agreement, the computational method remains limited due to the Hilbert space truncation, which prevents it from capturing states with high Fock-state populations. This highlights the power of the employed analytical tool. Moreover, we recall that in the limits $\eta\to0$ and $\eta\to1$, the state vectors $\ket{m}$ and $\hat{S}_{\xi_\text{in}}\ket{0}$ are being projected onto $\ket{n}$ by PNRD, respectively, which we now compare to Figs.\  \ref{fig:stellar2_theo} and \ref{fig:stellar3_theo}. Indeed, this behavior matches the limits of the success probability $P_\text{PC}\to|\braket{n}{m}|^2=\delta_{m,n}$ and $P_\text{PC}\to|\bra{n}\hat{S}_{\xi_\text{in}}\ket{0}|^2$ for $\eta\to0$ and $\eta\to1$, respectively. In the cases of $P_\text{PC}\neq0$, we can further read off the output state $\ket{\psi_\text{out}}\to\hat{S}_{\xi_\text{in}}\ket{0}$ and $\ket{\psi_\text{out}}\to\ket{m}$ for $\eta\to0$ and $\eta\to1$, respectively. Note that the condition $\Fcal_N>1/2$ is not fulfilled for the limit $\eta\to1$ in the cases of $m=3,n=0$ and $m=2,n=0$, where the output state $\ket{\psi_\text{out}}\to\ket{2},\ket{3}$ is not an optimal approximation to any cat state.


To further estimate the quality of the output state, we focus on two values: the direction of the cat squeezing and the success probability. Note that $|\alpha_\text{cat}|$, $|\xi_\text{cat}|$, and $\Fcal_N$ are less suitable for this purpose since they are directly related, as illustrated in Fig.\  \ref{fig:opt_state}. To assess the cat squeezing, Schlegel et al.\  established how cat amplitude and cat squeezing need to be directed in phase space to guarantee error resilience of cat coding: $\alpha_\text{cat}=|\alpha_\text{cat}|e^{i\theta}$ and $\xi_\text{cat}=|\xi_\text{cat}|e^{2i\theta}$ with some complex phase $\theta$ \cite{Schlegel_2022}. In our case, we can assume that cat squeezing is real and cat amplitude is either real or purely imaginary,  as derived in Eq.\  \eqref{eq:xi_cat} and formerly. This corresponds to $\theta=0,\frac{\pi}{2}$, such that the cases $\alpha_\text{cat}\in\R,\xi_\text{cat}\geq 0$ and $\alpha_\text{cat}\in i\R,\xi_\text{cat}\leq 0$ remain, respectively. 
In Figs.\  \ref{fig:stellar2_theo} and \ref{fig:stellar3_theo}, this means that the line styles of $\alpha_\text{cat}$ and $\xi_\text{cat}$ must coincide. This is fulfilled in the cases
\begin{align}\label{eq:(m,n)_correct_squ}
    (m,n)= (1,1),\, (1,2),\, (2,1)
\end{align}
and partially in (2,0), (3,0).

Higher stellar rank protocols, with $N\geq4$, can be further investigated. However, we anticipate that Eq.\  \eqref{eq:coef_coinc} becomes harder to fulfill since the number of conditions in Eq.\  \eqref{eq:coef_coinc} increases while the number of parameters $\eta,\xi_\text{in}$ remains the same. In other words, the dimension of the stellar rank $N$ space increases and conversely, the dimension of $\eta,\xi_\text{in}$ with optimal output state presumably decreases. For $N=4,5$, the range of $\eta$ in Fig.\  \ref{fig:stellar2_theo} shrinks from intervals to isolated points, i.e., $\eta$ and $\xi_\text{in}$ are strictly correlated. For $N=6,7$, this correlation even reduces to isolated values of $\eta,\xi_\text{in}$. 
In addition, we stress that higher stellar rank protocols correlate with higher input Fock states, which are challenging to generate in practice. Nonetheless, high photon number subtractions can be achieved at the cost of success probabilities and PC with $N>3$ is expected to become increasingly relevant with the improvement of experimental optical resources.

Finally, we emphasize that no other setup using the same number of photon additions and subtractions can exceed the achieved fidelity using PC, including higher number of modes.
To assess cat state approximations of catalysis output states, different metrics have been used so far. Hanamura et al. expressed the output core state in terms of control parameters \cite{Hanamura2026BeyondStellarRank} and employed the $x^2$-squeezing for optimization \cite{Kuchar2025nonlinearsqueezing}.
Crescimanna et al.\ investigated multimode PC schemes \cite{Crescimanna2024SeedGBSwithFock} and optimized the sum of fidelity and success probability $\Fcal + P_\text{PC}$ to simultaneously improve both quantities.
In contrast, we focus on fidelity and success probability successively: in Sec.\  \ref{s:Application to cat states}, the fidelity has already been optimized among states of bounded stellar rank, yielding the stellar fidelity $\Fcal_N$, which allowed us to set the stellar state vector $\ket{\psi_N}$ as target state and to derive optimal setup parameters. Among the solutions saturating the stellar fidelity, we will next optimize the success probability in the course of comparing PC to GBS-like protocols, which is discussed in the subsequent section.

\begin{figure}
  \centering
  \includegraphics[width=0.5\linewidth]{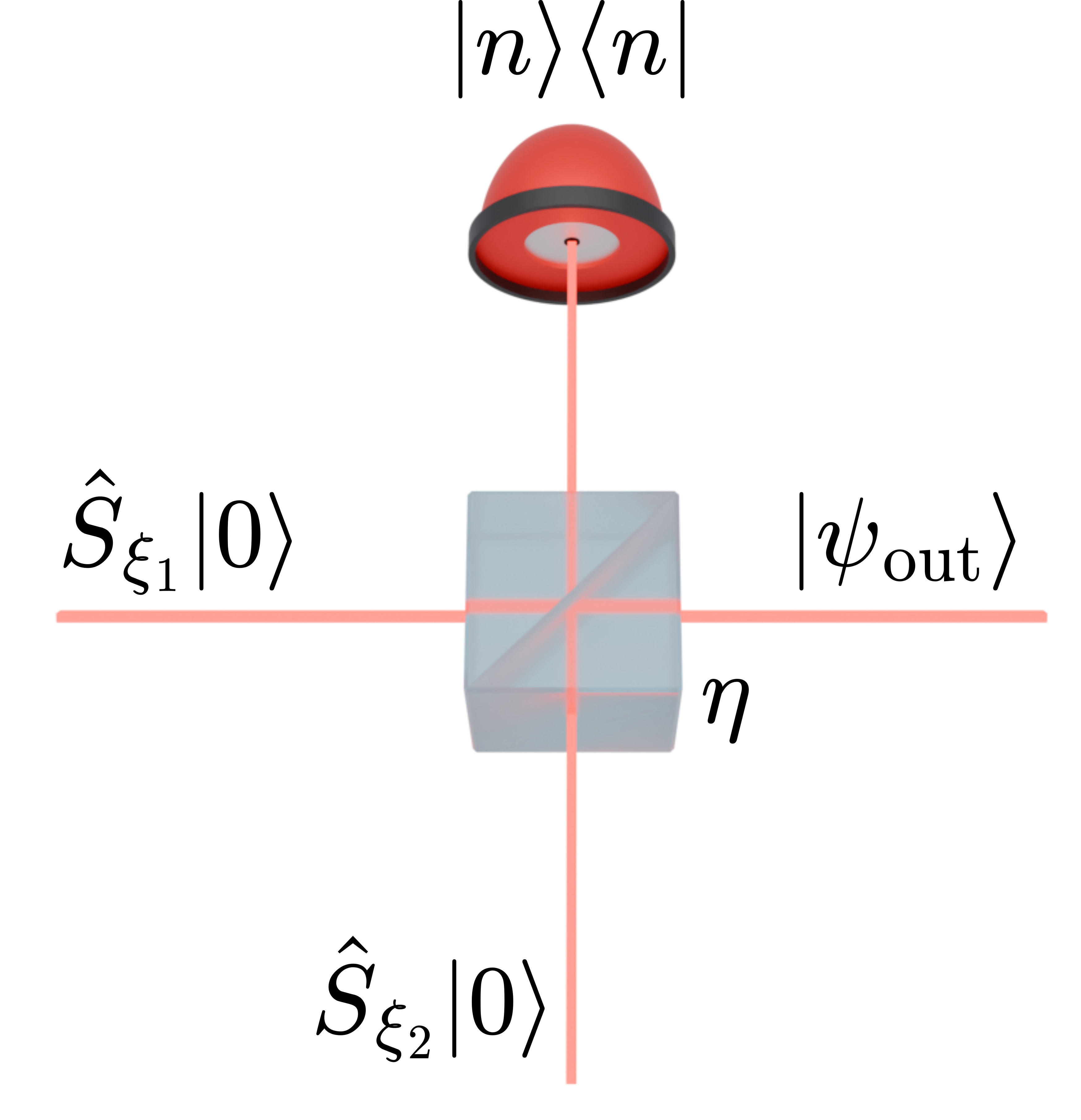}
  \caption{\emph{Gaussian-Boson-sampling}-like (GBS-like) scheme comprising a beam splitter with transmissivity $\eta$, two squeezed states $\hat{S}_{\xi_1}\ket{0}$, $\hat{S}_{\xi_2}\ket{0}$ as input states as well as a PNRD performed at one output port, i.e., a projection onto the Fock state vector $\ket{n}$. The remaining output port yields the state vector $\ket{\psi_\text{out}}$.}
\label{fig:GBS_setup}
\end{figure}

\section{Comparison to GBS-like scheme}\label{sec:comp_GBS}

To assess the success probability $P_\text{PC}$, given by the norm of the unnormalized output state in Eq.\  \eqref{eq:output_unnorm}, we compare PC with a two-mode GBS-like scheme, represented in 
Fig.\  \ref{fig:GBS_setup}. For this setting, Su et al.\  \cite{su2019pra} provide maximum achievable fidelities between the output state and squeezed cat states for fixed PNRD results $n=2,3$. Since the input states are Gaussian, the output states have stellar ranks $N=2,3$. Although the stellar rank had not been developed at this time, the fidelities in Ref.\ \cite{su2019pra} match the stellar fidelities $\Fcal_N$ in Fig.\  \ref{fig:opt_state}. Since PC and the GBS-like scheme both provide optimal stellar rank approximations, only the success probability remains for comparison. As suggested in Eq.\  \eqref{eq:(m,n)_correct_squ}, we investigate $(m,n)=$ (1,1), (1,2), (2,1) and compare them to the GBS-like scheme with the same stellar rank, i.e., $n=2$ and $n=3$, respectively. In particular, this covers examples for even and odd cat parities.
Analytical expressions of the success probability for two-mode GBS-like schemes are provided by Fiurášek \cite{Fiurasek2026maxprobability}.

Su et al.\  provide fidelities for various cat amplitudes $\alpha_\text{cat}$. Hence, we compare $P_\text{PC}$ and $P_\text{GBS}$ as a function of $\alpha_\text{cat}$ (see Fig.\  \ref{fig:Psucc_comp}). Efficient computation of $P_\text{PC}$ is provided in Appendix \ref{sec:appendix_Fock}. For PC, the values $(\alpha_\text{cat},P_\text{PC})$ are displayed as curves that are obtained from varying the transmissivity $\eta$. As input squeezing, we consider $\xi_\text{in}=5,10,15$ dB, where the latter is the largest squeezing that has been experimentally measured so far \cite{PhysRevLett.117.110801}. The case of $m=1,n=2$ gives rise to two branches, which originate from the gap at $\eta\approx 0.69$ as depicted in Fig.\  \ref{fig:stellar3_theo}. We infer from Fig.\  \ref{fig:Psucc_comp} that PC outperforms the GBS-like scheme for a broad range of $\alpha_\text{cat}$.

\begin{figure*}
\centering
  \includegraphics[width=0.78\linewidth]{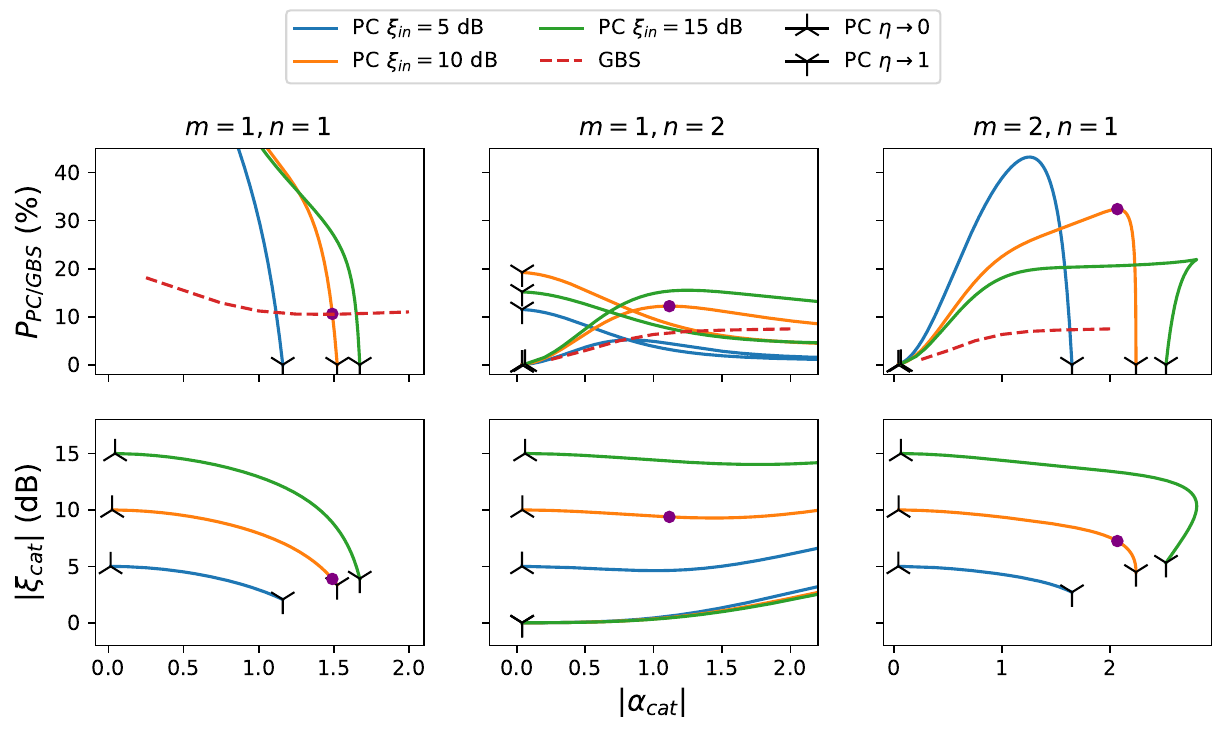}
  \caption{Dependence of the success probabilities $P_\text{PC}$ and $P_\text{GBS}$ achieved by PC and the GBS-like scheme, respectively, as well as the cat squeezing $\xi_\text{cat}$ achieved by PC on the cat amplitude $\alpha_\text{cat}$. Bold lines display $P_\text{PC}$ and $\xi_\text{cat}$ with different colors labeling the input squeezing $\xi_\text{in}=5,10,15$ dB. Dashed lines illustrate $P_\text{GBS}$, which stem from Su et al.\  \cite{su2019pra}. In the case of PC, the lines originate from varying the transmissivity $\eta$ and the limits $\eta\to 0,1$ are presented with respective markers. The cases $(m=1,n=1)$ and $(m=1,n=2),(m=2,n=1)$ for PC correspond to $n=2$ and $n=3$ for the GBS-like scheme, where the stellar rank is 2 and 3, respectively. The examples in Tab.\ \ref{tab:examples} are indicated as purple dots.}
\label{fig:Psucc_comp}
\end{figure*}

Let us start with some special cases: For $m=1,n=1$, the limit $\alpha_\text{cat}\to0$ yields $P_\text{PC}\to |\braket{1}{1}|^2 =100\%$ in contrast to $P_\text{GBS}\to |\bra{2}\hat{S}_{\xi_\text{in}}\ket{0}|^2$. For $m=1,n=2$, the limit of $\alpha_\text{cat}\to0$ yields $P_\text{GBS}\to |\bra{3}\hat{S}_{\xi_\text{in}}\ket{0}|^2 =0\%$ since squeezed states have even Fock occupation. In contrast, $P_\text{PC}\to|\bra{2}\hat{S}_{\xi_\text{in}}\ket{0}|^2$ reaches up to $19.2\%$, obtained for $\xi_\text{in}\approx 10$ dB. Moreover, PC outperforms the GBS-like scheme already for input squeezing of $\xi_\text{in}=5$ dB, whereas Su et al.\  employ input squeezing $\xi_1,\xi_2$ up to $10-14$ dB \cite{su2019pra}, which is experimentally realizable but cumbersome. The most prominent result is achieved for $m=2,n=1$ and $\xi_\text{in}=5$ dB, where the success probability of PC reaches up to $P_\text{PC}=43.2\%$ at $|\alpha_\text{cat}|\approx1.252$.

For a fair comparison, however, we also need to discuss the generation efficiency of the input states. The GBS-like scheme solely relies on squeezed states as inputs that can be generated deterministically, in contrast to Fock states required for PC. It is important to emphasize this issue, since exotic inputs can easily generate exotic outputs but their origin equally needs to be accounted for. The generation of a Fock input $\ket{m}$ becomes harder for increasing $m$, i.e., the extraction efficiency rapidly decreases. The simplest case $m=0$, however, is a special case of the GBS-like scheme and will thus not provide any improvement. For $m=1$, quantum dots provide a generation efficiency of up to $87\%$ \cite{yang24phot,gines2022prl}, which is most suitable for PC. Indeed, even when taking $87\%$ into account in the cases $m=1,n=1$ and $m=1,n=2$, Figure  \ref{fig:Psucc_comp} shows that PC still outperforms the GBS-like scheme for a broad range of $\alpha_\text{cat}$. It is worth noting that encouraging results on the generation of indistinguishable photons from disparate sources, namely quantum dots and spontaneous down conversion based sources, have been reported \cite{polyakov2011coalescence,huber2017interfacing,paudel2019direct}.

We next discuss the output squeezing. For the GBS-like scheme, experimental parameters can always be chosen such that arbitrary cat squeezing is obtained  \cite{su2019pra}, even independently of the cat amplitude and the success probability. In contrast, PC would require squeezed Fock inputs to achieve arbitrary cat squeezing. This, however, would be at the expense of the generation efficiency. For the sake of completeness, Fig.\  \ref{fig:Psucc_comp} also illustrates the range of cat parameters $(\alpha_\text{cat},\xi_\text{cat})$ that PC achieves. This is likewise displayed as curves obtained from varying the transmissivity $\eta$. Once again, the case of $m=1,n=2$ gives rise to two branches originating from the gap at $\eta\approx 0.69$ as depicted in Fig.\  \ref{fig:stellar3_theo}. It can be inferred that a broad range of cat parameters $(\alpha_\text{cat},\xi_\text{cat})$ can be achieved by adjusting the input squeezing $\xi_\text{in}$.

Finally, we select examples for each case $(m,n)=$ (1,1), (1,2), (2,1). Input squeezing is set to be $\xi_\text{in}=10$ dB to guarantee realistic experimental implementation. Moreover, we require that PC outperforms the GBS-like scheme in terms of success probability, i.e.,
\begin{align}\label{eq:PCbeatsGBS}
    P_\text{PC}> P_\text{GBS} \,.
\end{align}
The examples selected in the following are presented in 
Tab.\ \ref{tab:examples}. For $m=1,n=1$, we infer from Fig.\  \ref{fig:Psucc_comp} that Eq.\  \eqref{eq:PCbeatsGBS} imposes an upper bound on the cat amplitude. We select the highest value as an example. For $m=1,n=2$, there are two branches under consideration. The branch pertaining to $\eta\to1$ exhibits a broad range to satisfy Eq.\  \eqref{eq:PCbeatsGBS}, particularly in the limit $\eta\to1$. This region, however, corresponds to vanishing cat squeezing as inferred from Fig.\  \ref{fig:Psucc_comp}. Since sufficient cat squeezing is valuable for cat coding \cite{Schlegel_2022}, we consider the branch pertaining to $\eta\to0$, which provides $\xi_\text{cat}\approx\xi_\text{in}$. In Fig.\ \ref{fig:Psucc_comp}, we show that $P_\text{PC}$ has a maximum, which we select as an example. For $m=2,n=1$, we also select the maximum of $P_\text{PC}$, see Fig.\  \ref{fig:Psucc_comp}.

\begin{table}
\centering
{\renewcommand{\arraystretch}{1.2}
\begin{tabular}{ccccccccc}
\hline\hline
$(m,n)$ && $\eta$ && $|\alpha_\text{cat}|$ && $|\xi_\text{cat}|$ && $P_\text{PC}$
\\ \hline
(1,1) && 0.909 && 1.490 && 3.883 dB && 10.6 \%
\\ \hline
(1,2) && 0.107 && 1.115 && 9.379 dB && 12.2 \%
\\ \hline
(2,1) && 0.561 && 2.067 && 7.238 dB && 32.4 \%
\\\hline\hline
\end{tabular}
}
\caption{Examples of cat state generation with PC and input squeezing $\xi_\text{in}=10$ dB as indicated in Fig.\  \ref{fig:Psucc_comp}. For $m=1,n=1$, we select the highest $|\alpha_\text{cat}|$, such that $P_\text{PC}> P_\text{GBS}$. For $m=1,n=2$, we select the maximum of $P_\text{PC}$ of the branch pertaining to $\eta\to0$. For $m=2,n=1$, we select the maximum of $P_\text{PC}$.}
\label{tab:examples}
\end{table}

\section{Generation of squeezed 3-Fock states}
\label{sec:squeezed_fock}

Beyond squeezed cat states, squeezed Fock states $\hat S_\xi\ket{n}$ are equally of considerable interest for applications such as quantum error correction. Bashmakova et al.\   \cite{Bashmakova2025squeezedFockQEC} have defined a code with logical state vectors
\begin{align}\label{eq:sq_fock_code}
    \ket{0_L}= \hat S_\xi\ket{n} \,,\quad
    \ket{1_L}= \hat S_{-\xi}\ket{n},
\end{align}
and shown for $n=2$ that a finite squeezing parameter $\xi$ exists, such that $\ket{0_L},\ket{1_L}$ are orthogonal. In this section, we explore the possibility of generating squeezed Fock states using PC and of satisfying the orthogonality of the logical states in Eq.\  \eqref{eq:sq_fock_code}.

Initially, we draw attention to the prominent behavior of the case $m=1,n=2$ in Fig.\  \ref{fig:stellar3_theo}. The cat amplitude seems to diverge for non-trivial transmissivity $\eta$ and the coherent components change their orientation in phase space. This suggests that the coherent components merge and form a rotationally symmetric state up to squeezing, i.e., a squeezed Fock state. Provided that the stellar rank is preserved in this limit, we anticipate that $\ket{\psi_\text{out}}=\hat{S}_{\xi_\text{out}}\ket{3}$ is generated. To corroborate this intuition, we consider the Fock coefficients of the output core state. Since the core state vector $\ket{\phi_\text{out}}$ that defines the output state vector $\ket{\psi_\text{out}}$ (as described in Sec.\ \ref{sec:cat_approx}) has odd Fock occupation up to $3$, it remains to investigate whether its first Fock coefficient vanishes. 

\begin{figure}
\centering
\includegraphics[width=0.8\linewidth]{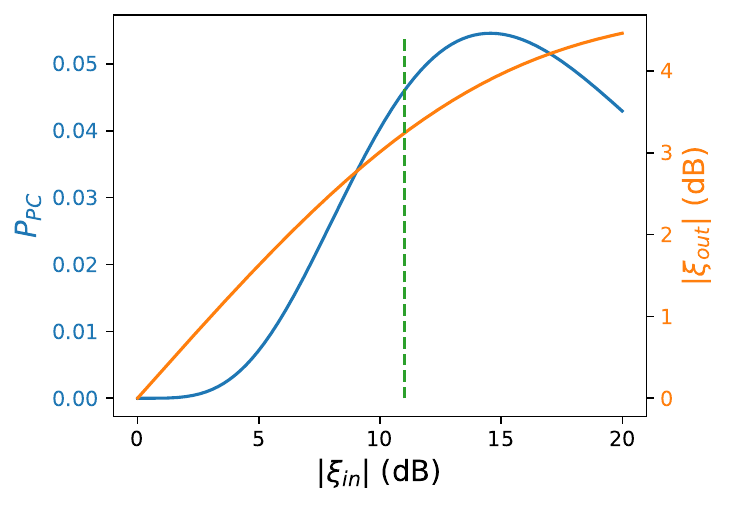}
  \caption{Dependence of the success probability $P_\text{PC}$ to generate $\ket{\psi_\text{out}}=\hat{S}_{\xi_\text{out}}\ket{3}$ and its output squeezing $\xi_\text{out}$ on the input squeezing $\xi_\text{in}$. The PC parameters are $m=1,n=2$ and the transmissivity $\eta$ obeys Eq.\  \eqref{eq:3fock_rel}. The green dashed line marks the input squeezing suitable to generate orthogonal squeezed Fock states.}
\label{fig:3Fock_Psucc_squeez}
\end{figure}

For the sake of simplicity, we again consider $\xi_\text{in}\in\R$. Based on Eq.\  \eqref{eq:output_Fock_eff} in Appendix \ref{sec:appendix_Fock}, we obtain
\begin{align}
    &\langle{1}|{\tilde\phi_\text{out}}\rangle
    = \sqrt{\frac{2\eta c_\text{out}^7}{c_\text{in}}} \left(\left[t_\text{in}(1-\eta)\right]^2 + \frac{3}{2}\eta -1\right)
\end{align}
for the unnormalized output core state vector $|\tilde\phi_\text{out}\rangle$, where $t_j = \tanh{\xi_j}$ and $c_j=\cosh{\xi_j}$ for $j= \text{in},\text{out} $.
Setting the parenthetic term to zero and solving the equation yields
\begin{align}\label{eq:3fock_rel}
    \eta = 1-\frac{3}{4t_\text{in}^2}\big(1-\sqrt{1-\frac{8}{9}t_\text{in}^2}\big) \,.
\end{align}
Hence, $\ket{\psi_\text{out}}=\hat{S}_{\xi_\text{out}}\ket{3}$ if Eq.\  \eqref{eq:3fock_rel} is fulfilled. Since $-1<t_\text{in}<1$ by definition, we obtain that $\frac{1}{2}<\eta\leq\frac{2}{3}$. In particular, Eq.\  \eqref{eq:3fock_rel} allows any value for $t_\text{in}$. This reveals that the generation of squeezed Fock states is possible for any input squeezing $\xi_\text{in}$. The output squeezing $\xi_\text{out}$, i.e., $t_\text{out}=\tanh{\xi_\text{out}}$, is determined by
\begin{align}\label{eq:3fock_tout}
    t_\text{out}=(1-\eta)t_\text{in} = \frac{3}{4t_\text{in}}\big(1-\sqrt{1-\frac{8}{9}t_\text{in}^2}\big) \,,
\end{align}
using Eq.\  \eqref{eq:3fock_rel}. Hence, it is limited by $|t_\text{out}|< \frac{1}{2}$, i.e., $|\xi_\text{out}|< 4.77$ dB. Based on Eq.\  \eqref{eq:3fock_tout}, the output squeezing is displayed in Fig.\  \ref{fig:3Fock_Psucc_squeez}.

Following Bashmakova et al.\  \cite{Bashmakova2025squeezedFockQEC}, we next address the orthogonality of the logical states in Eq.\  \eqref{eq:sq_fock_code} for $n=3$. To this end, we compute
\begin{align}
    \langle{0_L}|{1_L}\rangle = \bra{3}\hat{S}_{2\xi}\ket{3} =
    \big(1-\frac{5}{2}\tau^2\big)(1-\tau^2)^{3/4}
\end{align}
with $\tau=\tanh(2\xi)$, which vanishes for $\xi=\pm3.24$ dB. This lies within the range of achievable $\xi_\text{out}$. The corresponding input squeezing amounts to $|\xi_\text{in}|= 11.0$ dB, which is determined by Eq.\  \eqref{eq:3fock_tout} and also marked in Fig.\  \ref{fig:3Fock_Psucc_squeez}. Hence, PC can be used to generate orthogonal squeezed Fock states serving as logical states as in Eq.\  \eqref{eq:sq_fock_code}.

For completeness, we also discuss the success probability. If Eq.\  \eqref{eq:3fock_rel} is fulfilled, we obtain $P_\text{PC}=|\langle{3}|{\tilde\phi_\text{out}}\rangle |^2$. This simplifies to
\begin{align}
    P_\text{PC}= 3\frac{t_\text{out}^2\sqrt{1-t_\text{in}^2}}{t_\text{in}\sqrt{1-t_\text{out}^2}^7}(t_\text{in}-t_\text{out})^3 \,,
\end{align}
where $t_\text{out}$ is defined in Eq.\  \eqref{eq:3fock_tout} and depends on $t_\text{in}$. The success probability is also depicted in 
Fig.\ \ref{fig:3Fock_Psucc_squeez}. It can be inferred that the success probability reaches up to $P_\text{PC}=5.46\%$ at finite input squeezing $\xi_\text{in}=14.56$ dB and vanishes in the limit cases $\xi_\text{in}\to0,\infty$. The success probability to generate orthogonal squeezed Fock states, i.e., $|\xi_\text{in}|= 11.0$ dB, is $P_\text{PC}=4.59\%$.

\section{Loss analysis}
\label{sec:losses}

So far we have developed the PC protocol considering only pure states. In this section we analyze the mixed state scenario by exploring the effect of losses in the generation of squeezed cat states. Incorporating realistic noise models is essential to accurately predict fidelity degradation and to optimize parameters for feasible, high-quality state preparation. Common experimental imperfections, including nonideal state generation, propagation losses, photon-number-resolving detector inefficiencies, and mode mismatch, have been theoretically modeled~\cite{SUZUKI2006analysis,eaton2019non,dakna1998quantum, Arzani2019high, Datta:PRA:2025, Yoshikawa2017purification,Song2023entanglement} and show reasonable agreement with experimental results~\cite{Magana-Loaiza2019multiphoton,Zo2024entanglement}.

\begin{figure}[t!]
    \centering
    \includegraphics[width=.33\textwidth]{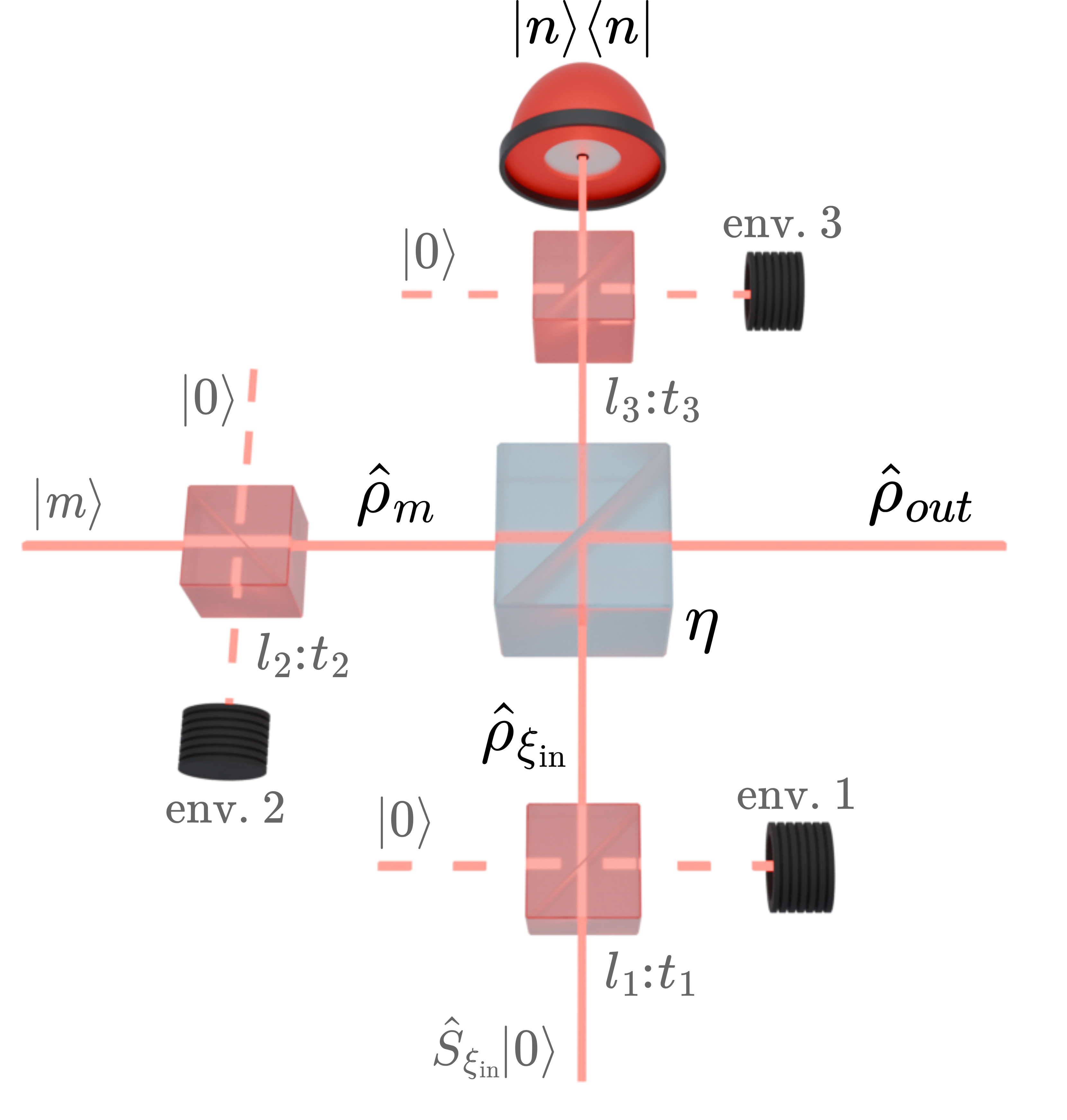}
    \caption{Loss modeling in PC for an input squeezed vacuum state vector $\hat{S}_{\xi_\text{in}}\ket{0}$ and a Fock state vector $\ket{m}$, using virtual beam splitters. The respective input states are now described by the density matrices $\hat{\rho}_{\xi_\text{in}}$ and $\hat{\rho}_m$. The detection efficiency of the heralding arm is modeled in an analogous way. Here, $l_j$ ($0 \leq l_j \leq 1$) denotes the loss parameter of the loss channel $j=1,2,3$, where the transmissivity $t_j$ of the virtual beam splitter satisfies $|t_j|^2+|l_j|^2=1$. PNRD of $n$ photons heralds the output state $\hat{\rho}_{\mathrm{out}}$.
}
    \label{fig: loss_Scheme}
\end{figure}

A standard method for incorporating optical losses is through a virtual beam splitter model that couples optical modes to a vacuum auxiliary. Tracing out the vacuum mode (i.e., performing a partial trace over the auxiliary) transforms a pure input into a mixed state, effectively capturing loss effects quantified by the splitting parameter~\cite{eaton2019non, Song2023entanglement}. Building on this framework, we model losses in both the input squeezed vacuum and the auxiliary Fock state, as well as in the heralding PNRD operation, as illustrated in Fig.~\ref{fig: loss_Scheme}. Despite the simplicity of the theoretical model, a full loss characterization can be highly demanding, even in protocols involving only two modes as under investigation here.

Each loss channel $j=1,2,3$ is characterized by a loss parameter $l_j$ ($0 \leq l_j \leq 1$) and represented by a virtual beam splitter with transmissivity $t_j$ satisfying $|t_j|^2+|l_j|^2=1$, 
which couples the mode to a vacuum environment auxiliary that is subsequently traced out. Specifically, loss in the squeezed vacuum state vector $\hat{S}_{\xi_\text{in}}\ket{0}$ accounts for practical effects on the generation and propagation of such a non-classical resource. Typically, such states are generated via non-linear interactions and are unavoidably affected by spurious losses in the non-linear media, pump noise, and propagation losses before entering the catalysis stage~\cite{Venneberg2025bright, Houde2022waveguided}. Similarly, the auxiliary Fock state vector $\ket{m}$, which can be generated by heralding from two-mode squeezed states or quantum dot sources, is subject to mode impurity and generation inefficiency~\cite{Konrad2006heralded, cooper2013experimental, Fadrny2024experimental}. Finally, photon-number-resolving detectors are subject to finite quantum efficiency, dark counts, and limited photon number resolution, all of which degrade measurement fidelity~\cite{Calkins2013photon, Hummatov2023fast}. Heralding on $n$ photons projects the output state into $\hat{\rho}_{\mathrm{out}}$. The described loss sources are respectively denoted by $1$, $2$, and $3$, as indicated in Fig.\ \ref{fig: loss_Scheme}.

Let us consider the loss operation on the state \(\hat{\rho}_A\), generally described as
\begin{align}
    \hat{\rho}_{A}=\sum_{M,M'=0}^\infty c_{M,M'}\ket{M}_A {\bra{M'}}_A
\end{align}
in the Fock basis $\ket{M}_{A},\ket{M'}_{A}$. Considering a vacuum auxiliary in the other input mode \(B\) of the virtual beam splitter, the joint input state is simply
\begin{align}
    \hat{\rho}_{A,B}=\hat{\rho}_{A} \otimes \ket{0}_B{\bra{0}}_B.
\end{align}
After passing through the lossy channel, the mixed state in mode \(A\) is obtained as
\begin{align}
    \hat{\rho}_{\mathrm{loss}}=\mathrm{Tr}_B\!\left(\hat{\rho}^{\prime}\right),
\end{align}
where the evolved state is 
\begin{align}
    \hat{\rho}^{\prime}=\hat{U}(t)\,\hat{\rho}_{AB}\,\hat{U}^{\dagger}(t),
\end{align}
and the beam splitter operation $\hat{U}(t)$ is described according to Eq.\  (\ref{eq:beamsplitter}).
The analytical expression for the state after passing through the lossy channel is
\begin{align}\label{eq:density_matrix_lossy}
    \hat{\rho}_{\mathrm{loss}}=\sum_{M,M'=0}^\infty \sum_{k=0}^{\min(M,M')} C_{M,M',k,T}\, \ket{M - k}_A {\bra{M' - k}}_A,
\end{align}
with coefficients
\begin{equation}
    C_{M,M',k,t} = c_{M,M'} \sqrt{\binom{M}{k} \binom{M'}{k}} \, t^{(M+M'-2k)/2} (1-t)^{k}.
\end{equation}
We employ Eq.~\eqref{eq:density_matrix_lossy} to describe the analytical form of the quantum state after propagation through the lossy channel. A detailed derivation is provided in Appendix~\ref{sec:appendix_a}.

Next, we consider the loss contributions to the catalysis protocol described in Sec.\ \ref{sec:catalysis}. Using Eq.~\eqref{eq:density_matrix_lossy}, the input squeezed state is now 
\begin{align}\label{eq:squeezed_lossy}
\begin{split}
\hat{\rho}_{\xi_\text{in}}=
\sum_{q,q'=0}^{\infty} \sum_{k_A=0}^{\min(2q,2q')} C^{\mathrm{sq}}_{q,q',k_A, t_1} \, \ket{2q - k_A}_A {\bra{2q' - k_A}}_A,
\end{split}
\end{align}
where the coefficients are
\begin{align}\label{eq:C_sq}
    C^{\mathrm{sq}}_{q,q',k_A, t_1}=c_{q,k_A,t_1}c^{*}_{q',k_A, t_1},
\end{align}
with
\begin{align}
    c_{q,k_A}=s_q \sqrt{\binom{2q}{k_A}}\Big(\sqrt{t_1}\Big)^{2q-k_A}\Big(i\sqrt{1-t_1}\Big)^{k_A},
\end{align}
and $s_q$ defined in Eq.~\eqref{eq:s_m}.
Similarly, utilizing Eq.~\eqref{eq:density_matrix_lossy}, the transformation of an $m$-photon Fock state under loss in mode $B$ results in the mixed state
\begin{align}
\hat{\rho}_m = \sum_{k_B=0}^{m} C^{\mathrm{Fock}}_{m, k_B, t_2} \, \ket{m - k_B}_B {\bra{m - k_B}}_B,
\end{align}
where 
\begin{align}\label{eq:C_fock}
C^{\mathrm{Fock}}_{m,k_B,t_2} = \binom{m}{l} t_2^{m - k_B} (1 - t_2)^{k_B}.
\end{align}
The losses on the squeezed and the Fock state inputs are respectively given by $l_1 = 1 - t_1$ and $l_2 = 1 - t_2$.

The catalysis protocol is completed after the Fock state projection results $n$ in one of the output modes with an overall detection efficiency modeled by $t_3$. In analogy to Eq.\  (\ref{eq:output_unnorm}), the unnormalized output state is given by
\begin{align}\label{eq:heralded_state}
    \hat{\rho}_{\rm{out}} = {\bra{n}} \mathrm{Tr}_{v} \left[  \hat{U}(t_3) \left( \hat{U}(\eta) \hat{\rho}_{\xi_\text{in}} \otimes \hat{\rho}_m \hat{U}(\eta)^{\dagger} \right)\hat{U}^\dagger(t_3) \right]\ket{n},
\end{align}
where $\mathrm{Tr}_{v}$ denotes the partial trace after coupling the vacuum mode through $\hat{U}(t_3)$ in the detection arm. The subscripts indicating the different optical modes were omitted for simplicity. The complete analytical expression of $\hat{\rho}_{\rm{out}}$ is given in Appendix~\ref{sec:appendix_b}.

\begin{figure*}[t!]
  \centering
  \includegraphics[width=.75\linewidth]{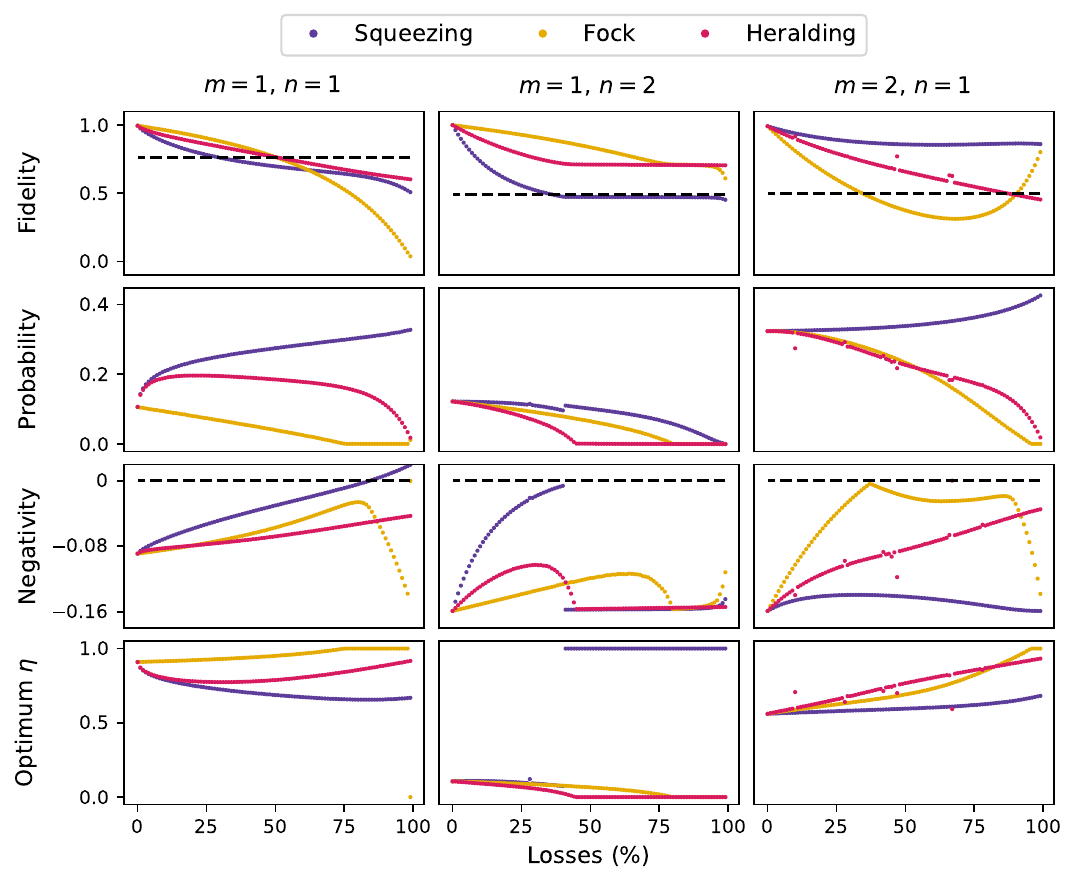}
  \caption{Effect of single channel losses to the output state $\hat{\rho}_{\rm{out}}$. The dependence of the fidelity, the success probability, the Wigner negativity, and the optimum transmissivity $\eta$ on a single loss parameter is displayed. Different line colors refer to the loss channels in each of the three PC arms that are illustrated in Fig.\ \ref{fig: loss_Scheme}: squeezing input (purple), Fock state input (yellow), and PNRD (red). The fidelity is calculated between $\hat{\rho}_{\rm{out}}$ and a fixed target state equivalent to the examples shown in Table \ref{tab:examples}, where the respective $(m,n)$ combinations are displayed in the columns. The fidelity is individually maximized in respect to the catalysis transmissivity $\eta$ for each loss configuration. The data set consists of $1000$ different transmissivity values $\eta$ and $100$ loss parameter values. In the plots illustrating the fidelity and Wigner negativity, the dashed lines refer to the Gaussian fidelity and zero, respectively.}
  \label{fig:HUBn2m1example-singlelosses-fixed_squeezing}
\end{figure*}

Following the method described in Sec.\ \ref{sec:catalysis}, we compare a list of target states with the explicitly computed output states $\hat{\rho}_{\rm{out}}(\eta,t_1,t_2,t_3)$ for fixed values of $\xi_{\rm{in}}$, $m$ and $n$, within a truncated Hilbert space of dimension $d=40$. We investigate the effects of single channel losses on PC protocols with $(m,n)=(1,1),(1,2),(2,1)$ and fixed input squeezing of $10$ dB. For each protocol, we explicitly compute the output state density matrices for $1000$ different PC transmissivity values ($\eta=\{0,0.001,\ldots ,1\}$), and for $100$ loss values in each of the protocol arms ($l_j = \{0,0.01,\ldots,1\}, j=\{1,2,3\}$). The targeted states correspond to the examples listed in Table \ref{tab:examples}. For each loss configuration, the fidelity between the output state and the corresponding target state is maximized by selecting the optimal transmissivity $\eta$. The effects of increasing losses can be seen in Fig.\ \ref{fig:HUBn2m1example-singlelosses-fixed_squeezing}, where we show the fidelity between the output state and the target state $\mathcal{F}=\bra{\psi_{\rm{cat}}}\hat{\rho}_{\rm{out}}(\eta)\ket{\psi_{\rm{cat}}}$, the success probability of detecting $n$ photons in the heralding operation, the minimum value of the Wigner function, and the transmissivity value $\eta$ that guarantees the maximum fidelity to the target state. The obtained fidelities are directly compared with the Gaussian fidelity $\Fcal_0$, defined as the maximum fidelity achievable by a Gaussian state approximating the target (unsqueezed) cat state, as shown in Fig.\  \ref{fig:opt_state}. We also note situations where the Wigner function attains only positive values, such that non-classicality is no longer guaranteed. Nevertheless, negativity can be preserved even in the presence of substantial losses with a non-trivial transmissivity result, i.e., $0<\eta<1$. From the plots, one can observe that the adopted fidelity maximization procedure often tends towards trivial solutions, as the overlap between the target state and the pure state of the arm unaffected by the single channel loss surpasses any other non-trivial output. 

Because the output states are sampled on discrete grids of both $\eta$ and loss values, isolated numerical outliers can appear in the optimized results. As discussed in Appendix \ref{sec:appendix_outliers}, these artifacts can be largely eliminated by combining multiple data sets with small offsets in the discretized parameters. The remaining outliers arise from the finite number of combined data sets and from near-degenerate fidelity values among competing solutions.

\begin{figure*}
  \centering
  \includegraphics[width=.75\linewidth]{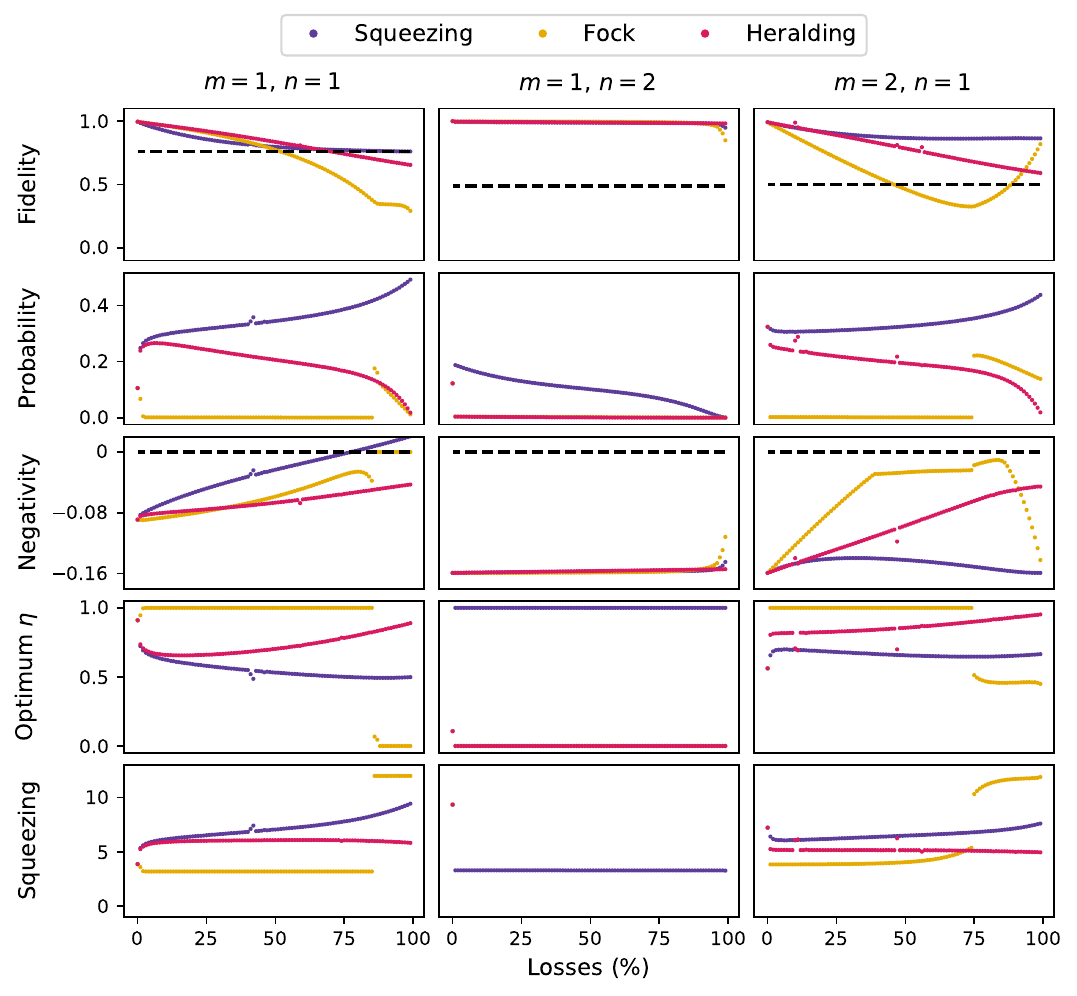}
  \caption{Similar to Fig.\  \ref{fig:HUBn2m1example-singlelosses-fixed_squeezing}, where the cat squeezing parameter of the target state is a further optimization variable.}
  \label{fig:HUBn2m1example-singlelosses-variable_squeezing}
\end{figure*}

One can further increase the fidelity of the generated state by leaving the target squeezing level as a free parameter, as effectively done in the previous sections. Figure \ref{fig:HUBn2m1example-singlelosses-variable_squeezing} shows the consequences of losses in this scenario, where catalysis protocols equivalent to the examples of Fig.\  \ref{fig:HUBn2m1example-singlelosses-fixed_squeezing} are chosen. In this scenario, the Gaussian fidelity is surpassed in a broader range of losses, where new optimal transmissivity coefficients are attributed to each loss value. This is especially appealing when considering squeezing as a simple Gaussian operation that can be applied after the PC. In practice, however, such an operation is not trivial and results in additional loss sources.

\begin{figure*}
  \centering
  \includegraphics[width=0.7\linewidth]{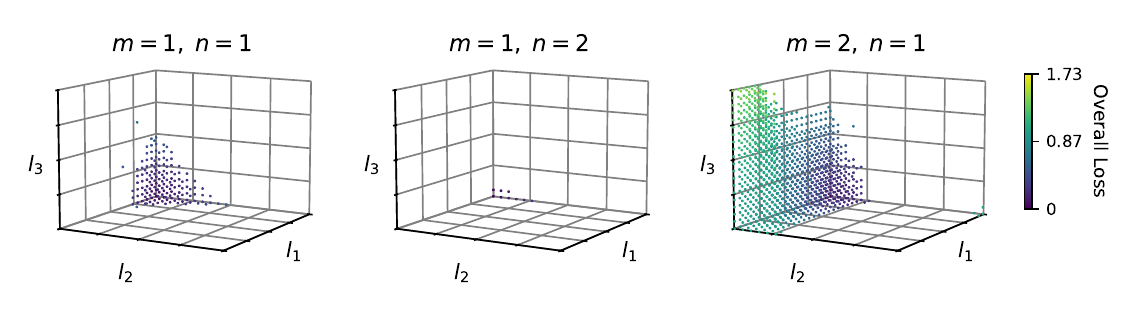}
  \caption{Multi channel loss analysis. Each point represents a combination of loss budgets $(l_1,l_2,l_3)$ for which the Gaussian fidelity with a fixed target state is surpassed. Catalysis and target state parameters as well as the optimization procedure, including fixed target squeezing, are equivalent to the single channel loss case in Fig.\  \ref{fig:HUBn2m1example-singlelosses-fixed_squeezing}. Each loss parameter pertains to a loss channel in each of the three PC arms that are illustrated in Fig.\ \ref{fig: loss_Scheme}. The displayed loss axes range from $0$ to $1$. The color-code represents the distance from the zero loss origin $d=\sqrt{l_1^2 + l_2^2 + l_3^2}$. }
  \label{fig:HUBn2m1example-multi_loss_fixed}
\end{figure*}

\begin{figure*}
  \centering
  \includegraphics[width=0.7\linewidth]{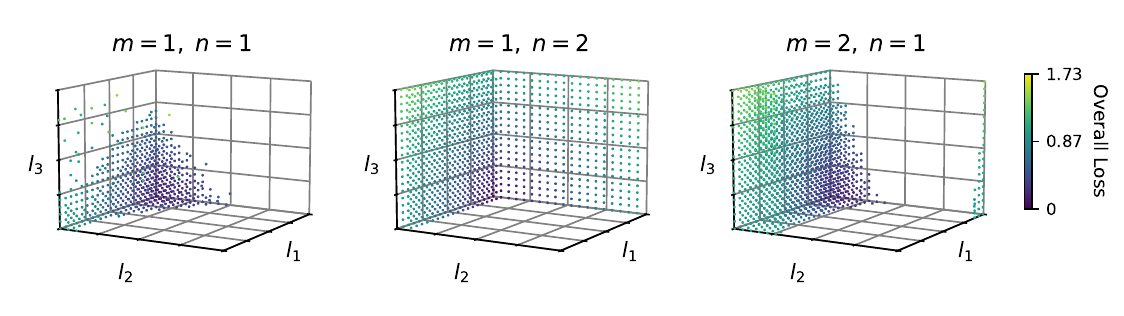}
  \caption{Similar to Fig.\  \ref{fig:HUBn2m1example-multi_loss_fixed}, where the cat squeezing parameter of the target state is a further optimization variable, equivalent to the single channel loss case in Fig.\  \ref{fig:HUBn2m1example-singlelosses-variable_squeezing}.}
  \label{fig:HUBn2m1example-multi_loss_variable}
\end{figure*}

\begin{figure*}[t!]
    \centering
    \includegraphics[width=0.7\textwidth]{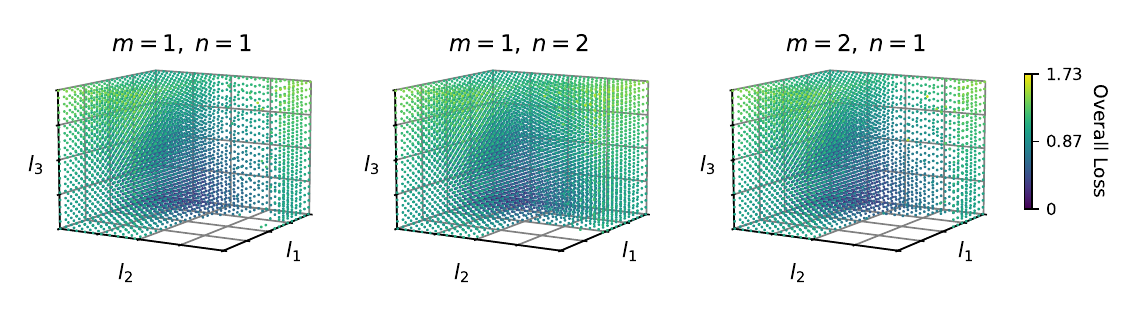}
    \caption{Similar to Fig.\  \ref{fig:HUBn2m1example-multi_loss_fixed}, where Wigner negativity is used as threshold for depicting loss budget values $(l_1,l_2,l_3)$.
    }
    \label{fig:HUBn2m1example-multi_loss-negativity}
\end{figure*}

An equivalent procedure is carried out in the multi-loss context. That is, for a given catalysis with input state vectors $\hat{S}_{\xi_{\text{in}}}\ket{0}$ and $\ket{m}$, a heralding on $\ket{n}$ photons and with respective losses $l_1, l_2$ and $l_3$, we determine the optimal transmissivity parameter $\eta$ such that $\hat{\rho}_{\text{out}}$ best approximates the respective target states of Table \ref{tab:examples}. Here, we evaluate the different loss combinations according to the Gaussian fidelity limit in both cases: fixed (Fig.\  \ref{fig:HUBn2m1example-multi_loss_fixed}) and variable (Fig.\  \ref{fig:HUBn2m1example-multi_loss_variable}) target squeezing. The density matrices of the output states $\hat{\rho}_{\rm{out}}$ are calculated for all combinations of $20$ evenly distributed loss values per channel $l_i=\{0, 0.05, \ldots,1 \}, i=\{1,2,3\}$ and for $100$ different transmissivity values $\eta=\{0,0.01,\ldots,1\}.$
In addition, we show the preservation of the Wigner negativity with the combined losses in Fig.\  \ref{fig:HUBn2m1example-multi_loss-negativity}.

Although the Gaussian fidelity is not always surpassed for the chosen target state, non-classicality of the output state can be preserved even for high loss combinations. This has also been observed in the single-channel loss analysis in Figs. \ref{fig:HUBn2m1example-singlelosses-fixed_squeezing} and \ref{fig:HUBn2m1example-singlelosses-variable_squeezing}. Hence, given a loss budget, one could still take advantage of the catalysis protocol in the generation of non-Gaussian states that are non-classical. These can be further used as resources in amplification schemes \cite{lund2004conditional,takeoka2007conditional,sychev2017enlargement} or cascaded breeding protocols \cite{eaton2019non,winnel2024deterministic,konno2024logical}. Finally, an alternative approach is considered in Appendix \ref{sec:appendix_max_fidelity}, where we fix the PC output and find the squeezed cat state that best approximates it.

\section{Conclusion and outlook}

In this work, we have comprehensively investigated the performance of a PC scheme for the preparation of cat states relevant for photonic quantum error correction. Specifically, we have
investigated the PC protocol as a means to generate squeezed Schrödinger cat states comprising a range of amplitudes and squeezing values. In addition to numerical characterizations on a truncated Fock basis, solutions within arbitrarily large Hilbert spaces have been found with the aid of the stellar rank formalism. Optimum protocols for the generation of the target states have been determined considering non-Gaussian outputs of up to stellar rank $N=3$, while acknowledging realistic experimental conditions. Moreover, we have derived a relation between the target cat state amplitude and the minimum stellar rank of the engineered state that is necessary to achieve different fidelity benchmarks, similar to the approximate stellar rank formalism \cite{Hahn_2026}.

Moreover, we have compared PC with two-mode GBS-like protocols \cite{su2019pra}, which have shown the advantage of deterministic sources of Fock states in breeding protocols. Considering state-of-the-art quantum dot systems, significant improvement in the success probability can be readily achieved. This further motivates the development of hybrid photonic technologies that interconnect disparate sources \cite{polyakov2011coalescence,huber2017interfacing,paudel2019direct}. 
In addition, we have found suitable PC parameters for the direct generation of squeezed Fock states. In particular, we have determined the feasibility of breeding orthogonal squeezed Fock states suitable for error correction protocols \cite{Bashmakova2025squeezedFockQEC}.

Working on a truncated Fock basis for direct numerical computation of the output density matrices, we have investigated the role of losses in the three resources necessary for PC, i.e., the squeezed state input, the Fock state input, and the photon number resolution measurement. We have determined the loss budgets where the approximation between the output state and a given target state is still advantageous in comparison to a trivial Gaussian approximation. Finally, we have computed the loss budgets that still provide a non-classical output through the preservation of the state's Wigner negativity. Despite the degradation of fidelity to targeted states in the presence of losses, we note that the output non-Gaussian states still represent valuable resources for amplification schemes  \cite{lund2004conditional,takeoka2007conditional,sychev2017enlargement} and iterative breeding protocols \cite{eaton2019non,winnel2024deterministic,konno2024logical}. We hope that the present work contributes to the body of work that bridges the gap between small-scale implementations and large-scale photonic preparations of resource states for quantum error correction \cite{MindTheGaps}.

Although the presented results align well with the limits of current experimental capabilities, this work can be extended to higher stellar rank outputs either by increasing the Fock number state input or the PNRD measurement result. Such a scenario could allow for the generation of different target states, possibly including the direct breeding of approximate GKP states. In addition, multi-mode schemes could benefit from Fock state resources for both single- and multi-mode non-Gaussian state engineering. A comparison between multi-mode GBS-like and hybrid setups in terms of success probability and loss resilience is clearly called for. Finally, we note that more exotic states, such as squeezed Schrödinger cats, squeezed Fock states, and photon added coherent states, can also be exploited in breeding protocols. However, the evaluation of  general optimized protocols under realistic considerations continues to be challenging and is expected to be further developed in future work.

\begin{acknowledgments}
J.K.N.\ and R.A.K.\ and J.E.\ 
acknowledge funding by the Senat Berlin via Berlin Quantum,
for this is a result of a joint node collaboration within Berlin.\ R.A.K.\ and O.B.\ acknowledge funding from the BMFTR PhoQuant project: 13N16105.\ A.M.D.\ and K.B.\ acknowledge financial support from the Leibniz Association through the Leibniz Collaborative Excellence Program (Project ID K266/2019, On-chip Laser-written Photonic Circuits for Classical and Quantum Applications (LAPTON)), and from the German Research Foundation (DFG) within the framework of Collaborative Research Center (CRC) 1375 (Project ID 398816777, Project A06). J.E.\ acknowledges funding via the BMFTR projects PhoQuant, PasQuops, and QPIC-1, the DFG (SPP 2514, BoLaCo), and the European Research Council (DebuQC).
\end{acknowledgments}

\appendix\markboth{Appendix}{Appendix}
\renewcommand{\thesection}{\Alph{section}}
\numberwithin{equation}{section}

\section{Efficient computation of stellar fidelities}
\label{sec:appendix_opt_fid}

In this appendix, we present an efficient way to compute the stellar fidelities $\Fcal_N$ of a cat state vector $\ket{\Ccal_{\alpha_\text{cat}}^\pm}$. To this end, we intend to avoid truncating $\hat{G}\ket{\Ccal_{\alpha_\text{cat}}^\pm}$ in the Fock basis. This preserves the nature of Gaussian unitary operators, such that the range of the Gaussian parameters is not restricted. In order to bring this into accordance with the Fock projection $\hat{\Pi}_N$, we recast the norm occurring in Eq.\ \eqref{eq:stellar_fid} as
\begin{align}\label{eq:Fopt_norm}
    \Vert \hat\Pi_N \hat{G}\ket{\Ccal_{\alpha_\text{cat}}^\pm} \Vert_2^2
    = \sum_{k=0}^N |\bra{k}\hat{G}\ket{\Ccal_{\alpha_\text{cat}}^\pm}|^2
\end{align}
and express the Fock state vector in terms of coherent state vectors as
\begin{align}\label{eq:Fock_deriv}
    \ket{k}=\frac{1}{\sqrt{k!}}\partial^k_{\beta}\big( e^{\frac{1}{2}|\beta|^2} \ket{\beta} \big|_{\beta=0} \,,
\end{align}
where we let $\beta\in\R$ in the following. This yields
\begin{align}\label{eq:Fopt_Fock}
    &\bra{k}\hat{G}\ket{\Ccal_{\alpha_\text{cat}}^\pm}
    \notag\\=& \frac{1}{\sqrt{k!}}\partial^k_{\beta}\big(e^{\frac{1}{2}|\beta|^2} \bra{\beta}\hat{G}\ket{\Ccal_{\alpha_\text{cat}}^\pm}\big|_{\beta=0}
    \notag\\ =& \frac{1}{\sqrt{k!\,C^\pm_{\alpha_\text{cat}}}} \sum_{s=0}^1 (\pm 1)^s \partial^k_{\beta}\big( e^{\frac{1}{2}|\beta|^2} \bra{\beta}\hat{G}\ket{(-1)^s\alpha_\text{cat}}\big|_{\beta=0}\,.
\end{align}
In what follows, we 
use standard techniques to compute the scalar product of Gaussian state vectors in the argument of the derivative \cite{MaRhodes1990GaussianStuff}. For $\hat{G}=\hat{S}_\xi\hat{D}_\alpha$, this gives
\begin{align}
    &e^{\frac{1}{2}|\beta|^2}\bra{\beta}\hat{G}\ket{\gamma}
    \\=&
    \frac{1}{\sqrt{c}} e^{-\frac{1}{2}(|\gamma|^2+|\alpha|^2)-\gamma\alpha^*}
    e^{\frac{1}{2}\big(t^*(\gamma+\alpha)^2-t\beta^2\big)-\frac{1}{c}\beta(\gamma+\alpha)} \,,
    \nonumber
\end{align}
where $t=({\xi}/{|\xi|})
\tanh{|\xi|}$, $c=\cosh|\xi|$, and $\gamma=(-1)^s\alpha_\text{cat}$. This facilitates the computation of Eq.\  \eqref{eq:Fopt_norm} to derivatives of a Gaussian function, which can be carried out efficiently using numerical tools.

Next, we address the Fock occupation of the optimal cat state approximation, given by the right-hand side of Eq.\  \eqref{eq:coef_coinc}. This is also covered by the preceding computation as
\begin{align}
    \braket{k}{\phi_N} = \bra{k}\hat\Pi_N \hat{G}_\text{opt}\ket{\Ccal_{\alpha_\text{cat}}^\pm}
    = \bra{k}\hat{G}_\text{opt}\ket{\Ccal_{\alpha_\text{cat}}^\pm}
\end{align}
for $k\leq N$, which can be computed efficiently using Eq.\  \eqref{eq:Fopt_Fock} and the optimal parameters $\alpha_\text{cat},\xi_\text{cat}$ obtained from computing $\Fcal_N$.

\section{Efficient computation of output state}
\label{sec:appendix_Fock}

In this appendix, we provide details of computing the decomposition in Eq.\  \eqref{eq:output_dec} of the output state. This covers the output squeezing $\xi_\text{out}$ and the Fock occupation of the output core state vector $\ket{\phi_\text{out}}$, which occurs on the left-hand side of Eq.\  \eqref{eq:coef_coinc}. In addition, we compute the success probability $P_\text{PC}$.

To begin with, we consider the stellar function of the unnormalized output state vector $|{\tilde\psi_\text{out}}\rangle$ as
\begin{align}
    F_\text{out}(\alpha)
    =& e^{\frac{1}{2}|\alpha|^2}\langle{\alpha^*}|{\tilde\psi_\text{out}}\rangle
    \notag\\=& e^{\frac{1}{2}|\alpha|^2} \big(\bra{\alpha^*}\otimes\bra{n}\big) \,\hat{U}(\eta)\, \big(\ket{m}\otimes\hat{S}_{\xi_\text{in}}\ket{0}\big)
    \notag\\=& \frac{1}{\sqrt{m!n!}} \partial^m_\beta\partial^n_\gamma \big( e^{\frac{1}{2}(|\alpha|^2+|\beta|^2+|\gamma|^2)}
    \notag\\&\times\big(\bra{\alpha^*}\otimes\bra{\gamma}\big) \, \hat{U}(\eta) \, \big(\ket{\beta}\otimes\hat{S}_{\xi_\text{in}}\ket{0}\big)\big|_{\beta,\gamma=0}
\end{align}
using Eq.\  \eqref{eq:Fock_deriv} along with $\beta,\gamma\in\R$. The 
argument of the derivative again yields a Gaussian function
\begin{align}
    &e^{\frac{1}{2}(|\alpha|^2+|\beta|^2+|\gamma|^2)} \big(\bra{\alpha^*}\otimes\bra{\gamma}\big) \, \hat{U}(\eta) \, \big(\ket{\beta}\otimes\hat{S}_{\xi_\text{in}}\ket{0}\big)
    \notag\\=& \frac{1}{\sqrt{c_\text{in}}} e^{\beta(\sqrt{\eta}\alpha+i\sqrt{1-\eta}\gamma)+\frac{1}{2} t_\text{in} (\sqrt{\eta}\gamma+i\sqrt{1-\eta}\alpha)^2} \,,
\end{align}
where 
\begin{align}
t_\text{in}=\frac{\xi_\text{in}}{|\xi_\text{in}|}\tanh{|\xi_\text{in}|}
\end{align}
and $c_\text{in}=\cosh{|\xi_\text{in}|}$. This serves to express the output state vector 
in terms of the stellar function as $|{\tilde\psi_\text{out}}\rangle = F_\text{out}(\hat{a}^\dagger)\ket{0}$. To derive the core state, we extract the $\alpha^2$-factor in $F_\text{out}(\alpha)$ to give the output squeezing $\hat{S}_{\xi_\text{out}}$. More precisely, we define
\begin{align}
    F_\text{out}(\alpha) = e^{\frac{1}{2}t_\text{out}\alpha^2} f_\text{out}(\alpha)\,,
\end{align}
where the output squeezing $\xi_\text{out}$ satisfies $t_\text{out}=t_\text{in}(1-\eta)$ for \begin{align}
t_\text{out}= \frac{\xi_\text{out}}{|\xi_\text{out}|}\tanh(|\xi_\text{out}|)
\end{align}
as stated in Eq.\  \eqref{eq:xi_out}. This yields
\begin{align}
    f_\text{out}(\alpha) = \frac{1}{\sqrt{{c_\text{in}m!n!}} }\partial^m_\beta\partial^n_\gamma \big(& e^{\frac{1}{2}t_\text{in}\eta\gamma^2 + i\sqrt{1-\eta}\beta\gamma}
    \notag\\\times &e^{(it_\text{in}\sqrt{1-\eta}\gamma+\beta)\sqrt{\eta}\alpha} \big|_{\beta,\gamma=0}. 
\end{align}
From this,
we obtain
\begin{align}
    |{\tilde\psi_\text{out}}\rangle =& f_\text{out}(\hat{a}^\dagger) e^{\frac{1}{2}t_\text{out}(\hat{a}^\dagger)^2} \ket{0}
    \notag\\=& \sqrt{c_\text{out}} f_\text{out}(\hat{a}^\dagger) \hat{S}_{\xi_\text{out}}\ket{0}
    \notag\\=& \hat{S}_{\xi_\text{out}} \sqrt{c_\text{out}} f_\text{out}(c_\text{out}\hat{a}^\dagger+s_\text{out}\hat{a})\ket{0}\,,
\end{align}
where $c_\text{out}=\cosh|\xi_\text{out}|$ and 
\begin{align}
s_\text{out}= \frac{\xi_\text{out}}{|\xi_\text{out}|}\sinh|\xi_\text{out}|. 
\end{align}
Using the Baker-Campbell-Hausdorff formula, we get 
\begin{align}
|{\tilde\psi_\text{out}}\rangle = \hat{S}_{\xi_\text{out}}|{\tilde\phi_\text{out}}\rangle 
\end{align}
with the core state vector
\begin{align}
    |{\tilde\phi_\text{out}}\rangle
    =& \sqrt{c_\text{out}} f_\text{out}(c_\text{out}\hat{a}^\dagger+s_\text{out}\hat{a})\ket{0}
    \notag\\=& \sqrt{\frac{c_\text{out}}{c_\text{in}m!n!}} \partial^m_\beta\partial^n_\gamma \big( e^{\frac{1}{2}t_\text{in}\eta\gamma^2 + i\sqrt{1-\eta}\beta\gamma}
    \notag\\&\times e^{\delta_{\beta,\gamma}(\hat{a}^\dagger+t_\text{out}\hat{a})} \ket{0} \big|_{\beta,\gamma=0}
    \notag\\=& \sqrt{\frac{c_\text{out}}{c_\text{in}m!n!}} \partial^m_\beta\partial^n_\gamma \big( e^{\frac{1}{2}t_\text{in}\eta\gamma^2 + i\sqrt{1-\eta}\beta\gamma}
    \notag\\&\times e^{\frac{1}{2}
    (|\delta_{\beta,\gamma}|^2+t_\text{out}\delta_{\beta,\gamma}^2)} \ket{\delta_{\beta,\gamma}} \big|_{\beta,\gamma=0} \,,
\end{align}
where we have abbreviated 
\begin{align}
\delta_{\beta,\gamma}=(it_\text{in}\sqrt{1-\eta}\gamma+\beta)\sqrt{\eta}c_\text{out}. 
\end{align}
For the sought-after Fock coefficients in Eq.\  \eqref{eq:coef_coinc}, we expand the coherent state vector $\ket{\delta_{\beta,\gamma}}$ in the Fock basis and obtain
\begin{equation}\label{eq:output_Fock_eff}
\begin{split}
    \langle{k}|{\tilde\phi_\text{out}}\rangle &= \sqrt{\frac{c_\text{out}}{c_\text{in}k!m!n!}} \\ & \times \partial^m_\beta\partial^n_\gamma \big( 
     \delta_{\beta,\gamma}^k e^{\frac{1}{2}t_\text{in}\eta\gamma^2 + i\sqrt{1-\eta}\beta\gamma + \frac{1}{2}
    t_\text{out}\delta_{\beta,\gamma}^2}  \big|_{\beta,\gamma=0} \,,
\end{split}
\end{equation}
which can be computed efficiently. Moreover, it serves to compute the success probability as
\begin{align}
    P_\text{PC} = \langle{\tilde\phi_\text{out}}|{\tilde\phi_\text{out}}\rangle = \sum_{k=0}^{m+n} \left|\langle{k}|{\tilde\phi_\text{out}}\rangle\right|^2 \,,
\end{align}
where we used that the core state vector $|{\tilde\phi_\text{out}}\rangle $ has Fock occupation up to $N=n+m$ at most, which is the stellar rank of $|{\tilde\psi_\text{out}}\rangle$.

\section{Density matrix transformation in presence of loss}
\label{sec:appendix_a}

In this appendix, we provide the derivation of Eq.~\eqref{eq:density_matrix_lossy} using the virtual beam splitter model. Consider Fock states with \(N\) photons in mode \(A\) and \(M\) photons in mode \(B\) incident on a beam splitter. Under the beam-splitting transformation, the Fock state vectors evolve as
\begin{widetext}
\begin{equation}\label{eq:beam_splitting_transformation}
\begin{split}
\ket{N}_{A}\otimes\ket{M}_{B} \rightarrow &
\frac{1}{\sqrt{N! M!}}
\sum_{k=0}^{N}\sum_{l=0}^{M} 
\binom{N}{k} \binom{M}{l} 
(\sqrt{t})^{N+l-k} \\
& \times (i\sqrt{1-t})^{M+k-l} 
\sqrt{(M+N-k-l)! (k+l)!} \,
\ket{M+N-k-l}_{A} \otimes \ket{k+l}_{B},
\end{split}
\end{equation}
\end{widetext}
where \(t\) is the transmissivity of 
the beam splitter.
For the specific 
case of 
a vacuum in mode \(B\), \(\ket{M}_A \otimes \ket{0}_B\) transforms as
\begin{widetext}
\begin{equation}\label{eq:Fock_basis_transformation}
\ket{M}_A \otimes \ket{0}_B \;\mapsto\; \sum_{k=0}^{M} \sqrt{\binom{M}{k}} \, t^{(M-k)/2} \, (i\sqrt{1-T})^k \, \ket{M-k}_A \otimes \ket{k}_B.
\end{equation}
Similarly, the corresponding transformation applies to the dual vector  ${\bra{M'}}_A\otimes{\bra{0}}_B$. The loss state $\hat{\rho}_{\mathrm{loss}}$ is then obtained by taking the partial trace over mode 
$B$, to get
\begin{equation}
\hat{\rho}_{\mathrm{loss}} = \sum_{M,M'} c_{M,M'} \sum_{k=0}^{\min(M,M')} 
\sqrt{\binom{M}{k}\binom{M'}{k}} \, T^{(M+M'-2k)/2} (1-T)^k \, \ket{M-k}_A {\bra{M'-k}}_A.
\end{equation}
\end{widetext}

\section{Analytical expression of the heralded state}
\label{sec:appendix_b}

In this appendix, we derive the analytical expression of the joint output states obtained when 
$\hat{\rho}^{(\mathrm{sq})}_A$ and $\hat{\rho}^{(\mathrm{Fock})}_B$ interfere at a beam splitter
with transmissivity $\eta$. Using Eq.~\eqref{eq:beam_splitting_transformation}, the joint output state 
$\hat{\rho}^{\prime}$ is
\begin{widetext}
\begin{align}
    \hat{\rho}^{\prime} &= \sum_{q,q'=0}^{\infty} \sum_{k_A=0}^{\min(2q,2q')} \sum_{k_B=0}^{m} \sum_{r=0}^{2q - k_A} \sum_{s=0}^{m - k_B} \sum_{r'=0}^{2q' - k_A} \sum_{s'=0}^{m - k_B}
    Q \ket{n_A}_A{\bra{n_A'}}_A
    \otimes \ket{n_B}_B{\bra{n_B'}}_B,
\end{align}
where
\begin{align}\label{eq:Q2}
    Q &= C^{\mathrm{sq}}_{q,q',k_A,t_1} C^{\mathrm{Fock}}_{m,k_B,t_2} 
    \frac{1}{(m-k_B)!\sqrt{(2q-k_A)!(2q'-k_A)!}}
    \binom{2q - k_A}{r} \binom{m - k_B}{s} \binom{2q' - k_A}{r'} \binom{m - k_B}{s'} \nonumber \\
    &\quad \times \big(\sqrt{\eta}\big)^{(2q - k_A) + s - r} (i\sqrt{1 - \eta})^{(m - k_B) + r - s} \, 
    \big(\sqrt{\eta}\big)^{(2q' - k_A) + s' - r'}  
    (-i\sqrt{1 - \eta})^{(m - k_B) + r' - s'}  \nonumber \\
    &\quad \times \sqrt{(2q - k_A + m - k_B - r - s)! \, (r + s)! \, (2q' - k_A + m - k_B - r' - s')! \, (r' + s')!},
\end{align}
with
\begin{align}
n_A &= 2q - k_A + m - k_B - r - s, \\
n_A' &= 2q' - k_A + m - k_B - r' - s', \\
n_B &= r+s, \\
n_B' &= r'+s'.
\end{align}
\end{widetext}
with $C^{\mathrm{sq}}_{q,q',k_A,t_1}$ and $C^{\mathrm{Fock}}_{m,k_B,t_2}$ defined in Eqs.~\eqref{eq:C_sq} and \eqref{eq:C_fock}, respectively.
We aim to herald \( n \) photons by PNRD in mode \( b \), 
using a detector with quantum efficiency \( t_3 \). The imperfection is modeled by a virtual BS with 
transmission coefficient \( t_3 \) placed before an ideal detector.
Using Eqs.~\eqref{eq:density_matrix_lossy} and \eqref{eq:heralded_state}, the heralded state for detecting \( n \) photons is
\begin{widetext}
\begin{align}
    \hat{\rho}_{\mathrm{out}} &= \sum_{q,q'=0}^{\infty} \sum_{k_A=0}^{\min(2q,2q')} \sum_{k_B=0}^{m} \sum_{r=0}^{2q - k_A} \sum_{s=0}^{m - k_B} \sum_{r'=0}^{2q' - k_A} \sum_{s'=0}^{m - k_B}
    Q
    \binom{n_B}{n_B-n}
    t_3^{n}
    (1-t_3)^{n_B-n}\ket{n_A}_A{\bra{n_A'}}_A,
\end{align}
\end{widetext}
where $Q$ is defined in Eq.~\eqref{eq:Q2}.

\section{Outliers evaluation}
\label{sec:appendix_outliers}

\begin{figure}
  \centering
  \includegraphics[width=.8\linewidth]{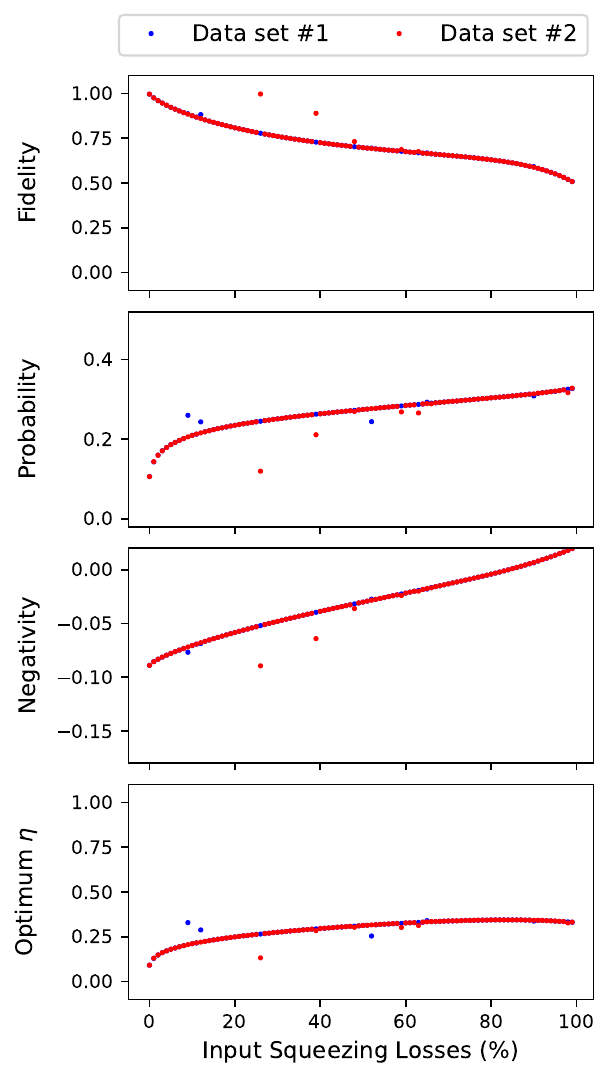}
  \caption{Numerical influence of different output state data sets to the single channel loss analysis. The dependence of the fidelity between the output state and a fixed target state, the success probability, the Wigner negativity of the output state, and the optimum transmissivity $\eta$ on the loss parameter $l_1$ is displayed. The corresponding loss channel acts on the squeezed input and is illustrated in Fig.\ \ref{fig: loss_Scheme}. The photon-catalysis protocol is configured with $(m,n)=(1,1)$ and compared with the corresponding target state listed in Table \ref{tab:examples}. For each loss value, the fidelity is maximized in respect to the catalysis transmissivity $\eta$ using two slightly offset discretizations of the parameter space, resulting in different distributions of numerical outliers. Different colors label the corresponding data sets (\#1) and (\#2).}
\label{fig:outliers_evaluation}
\end{figure}

As discussed in Section \ref{sec:losses}, spurious outliers arise in the optimization procedure due to the discrete sampling of the parameter space. As an example of these artifacts, we consider the $m=1,n=1$ protocol with fixed target squeezing and single loss applied solely to the input squeezed state. Two independent data sets of output states were generated using slightly different discretizations of the parameters: (\#1) $1000$ transmissivity values $\eta=\{0,0.001,\ldots,1\}$, and $100$ squeezing loss values $l_1=\{0, 0.05, \ldots,1 \}$; and (\#2)  $1000$ transmissivity values $\eta=\{0,0.001+0.0002,\ldots,1\}$, and the same set of squeezing loss values $l_1=\{0, 0.05, \ldots,1 \}$. The small offset on the data grid leads to a different distribution of numerical artifacts, as shown in 
Fig.\  \ref{fig:outliers_evaluation}. By combining the two data sets and removing isolated discrepant points, the influence of these discretization-induced errors is substantially reduced. The resulting filtered data are those presented in Fig.\ \ref{fig:HUBn2m1example-singlelosses-fixed_squeezing}.\\

\section{Maximum fidelity cat state for fixed catalysis parameters}
\label{sec:appendix_max_fidelity}

\begin{figure}
  \centering
  \includegraphics[width=.8\linewidth]{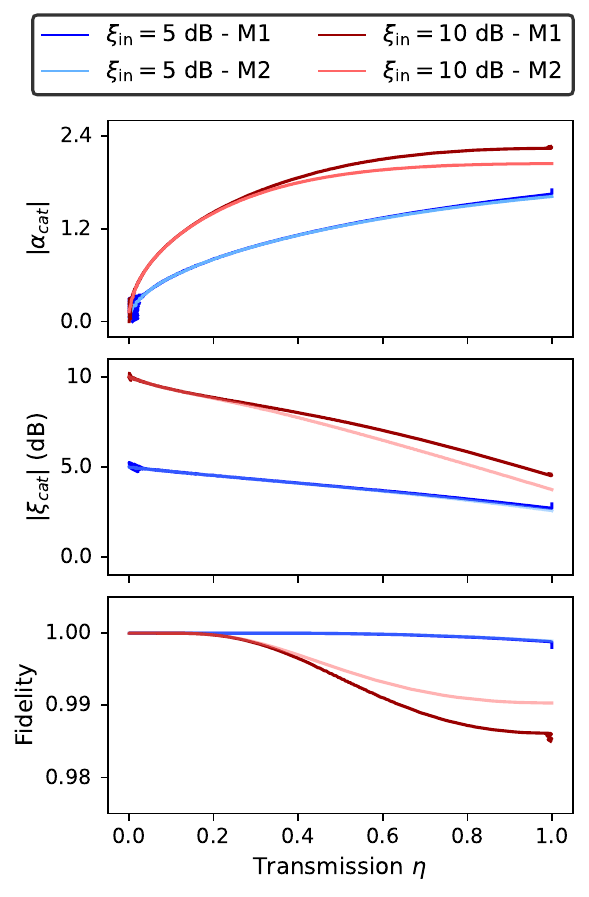}
  \caption{Comparison between the fidelity maximization of the output state to a fixed cat state (M1), treated in the main text, and to an arbitrary cat state (M2). The parameters of the considered protocols are $m=2$, $n=1$ and $|\xi_\text{in}|=5, 10$ dB.}
\label{fig:HUBn2m1example_M2}
\end{figure}

The quantum state engineering problem tackled in this work finds the optimal parameters for a catalysis protocol that generates the highest fidelity approximation to a given target output, hereby referred to as method 1 (M1). In this appendix, we invert the problem by fixing the catalysis parameters and finding the squeezed cat state that best approximates the output state (M2). That is, given the density matrix $\hat{\rho}_{\rm{out}}$, we determine $\alpha_{\rm{cat}}$ and $\xi_{\rm{cat}}$ that maximize the fidelity $\mathcal{F}=\bra{\psi_{\rm{cat}}(\alpha_{\text{cat}},\xi_{\rm{cat}})}\hat{\rho}_{\rm{out}}\ket{\psi_{\rm{cat}}(\alpha_{\text{cat}},\xi_{\rm{cat}})}$. The numerical procedure now consists of calculating the fidelity of a fixed output state with a list of squeezed cat states and selecting the maximum fidelity. A comparison between the pure state solutions from both methods considering the case with $m=2$ input photons, $10$ and $5$ dB of input vacuum squeezing and a PNRD result of $n=1$ is shown in Fig.\  \ref{fig:HUBn2m1example_M2}. These are the same examples explored in Sec.\ \ref{sec:catalysis}, shown in Fig.\  \ref{fig:HUBn2m1example}. At the cost of a more limiting amplitude range, the fidelity in describing squeezed cat states improves. One should note that the success probability is equivalent to the one found in the example shown in Fig.\  \ref{fig:HUBn2m1example}, as it is only dependent on the input states and the splitting parameter. This approach is especially relevant in situations where the described PC protocol is only a part of the full engineering process, as in iterative schemes \cite{eaton2019non,winnel2024deterministic,konno2024logical}.

\begin{figure}
  \centering
  \includegraphics[width=.8\linewidth]{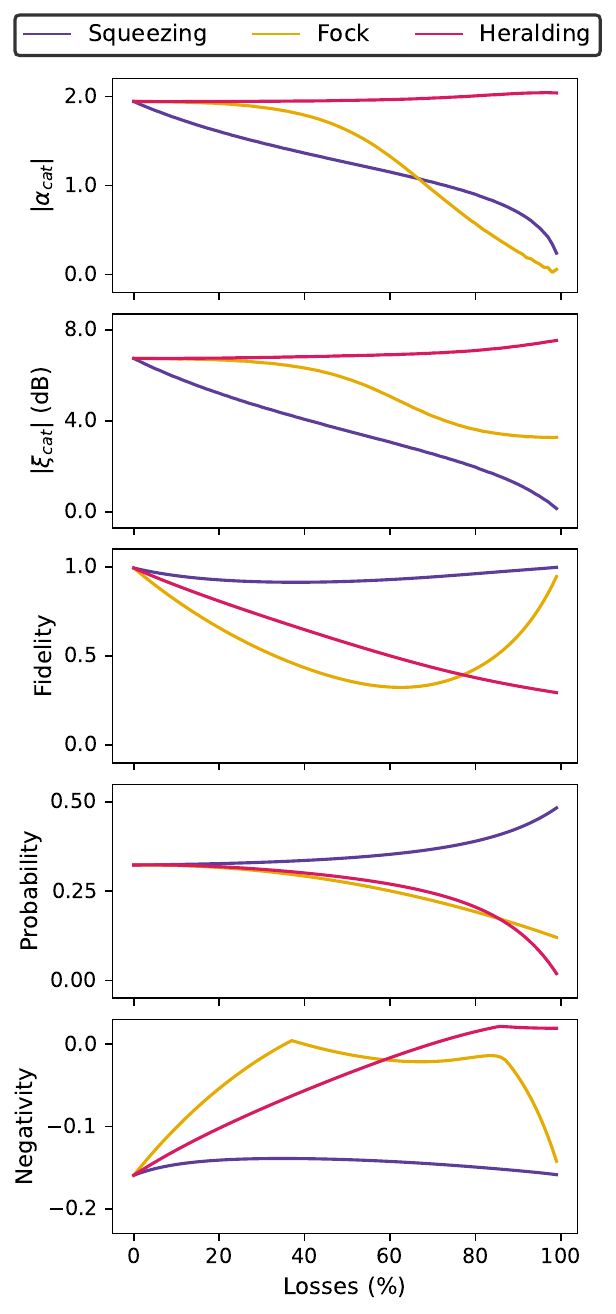}
  \caption{Fidelity maximization of the output state to an arbitrary cat state, as shown in M2 of 
  Fig.\  \ref{fig:HUBn2m1example_M2}, given a fixed catalysis scheme submitted to single channel losses. Catalysis parameters are equivalent to the lossless example $m=2,n=1$ in Table \ref{tab:examples}.}
\label{fig:HUBn2m1example_M2_single_losses}
\end{figure}

We further provide a single-channel loss analysis for the example ($2,1$) shown in Table \ref{tab:examples}, that is, with a catalysis protocol with fixed splitting parameter $\eta=0.561$. Figure  \ref{fig:HUBn2m1example_M2_single_losses} shows how losses affect the optimal obtained amplitude and squeezing, as well as the fidelity between the output state and the determined best squeezed cat approximation. The probability and Wigner negativity are also affected in a different way, as now the transmissivity parameter is maintained as a fixed parameter. The Gaussian fidelities are omitted from the picture for clarity, as now they will vary as a function of $|\alpha_\text{cat}|$. 
Extensions to multiple simultaneous loss channels can be carried out in a similar way as shown in Sec.\ \ref{sec:losses} and will not be explicitly shown here.

\bibliography{bib2}

\end{document}